\documentclass[preprint]{aastex631}
\shorttitle{Overview Orbit 10}
\shortauthors{Howard et al.}
\usepackage{gensymb}
\graphicspath{{./figures/}}
\newcommand{\R}{\textcolor{red}}
\newcommand{\B}{\textcolor{blue}}

\begin{document}

\title{Overview of the remote sensing observations from PSP solar encounter 10 with perihelion at 13.3~R$_\sun$}

\correspondingauthor{Russell Howard}
\email{russell.howard@jhuapl.edu}

\author[0000-0001-9027-8249]{Russell A. Howard}
\affiliation{Johns Hopkins University, 
Applied Physics Laboratory, 
Laurel, MD 20723, USA}

\author[0000-0001-8480-947X]{Guillermo Stenborg}
\affiliation{Johns Hopkins University, 
Applied Physics Laboratory, 
Laurel, MD 20723, USA}

\author[0000-0002-8164-5948]{Angelos Vourlidas}
\affiliation{Johns Hopkins University, 
Applied Physics Laboratory, 
Laurel, MD 20723, USA}

\author[0000-0002-8353-5865]{Brendan M. Gallagher}
\affiliation{Space Science Division, U.S. Naval Research Laboratory, 
Washington, DC 20375, USA}

\author [0000-0002-4459-7510]{Mark G. Linton}
\affiliation{Space Science Division, U.S. Naval Research Laboratory, 
Washington, DC 20375, USA}

\author [0000-0003-1377-6353]{Phillip Hess}
\affiliation{Space Science Division, U.S. Naval Research Laboratory, 
Washington, DC 20375, USA}

\author {Nathan B. Rich}
\affiliation{Space Science Division, U.S. Naval Research Laboratory, 
Washington, DC 20375, USA}

\author[0000-0002-5068-4637]{Paulett C. Liewer}
\affiliation{Jet Propulsion Laboratory, California Institute of Technology, Pasadena, CA 91125, USA}



\begin{abstract}

The closest perihelion pass of Parker Solar Probe (PSP), so far, occurred between 16 and 26 of November 2021 and reached $\sim$13.29~R$_\sun$ from Sun center. This pass resulted in very unique observations of the solar corona by the Wide-field Instrument for Solar PRobe (WISPR). WISPR observed at least ten CMEs, some of which were so close that the structures appear distorted.  All of the CMEs appeared to have a magnetic flux rope (MFR) structure and most were oriented such that the view was along the axis orientation, revealing very complex interiors. Two CMEs had a small MFR develop in the interior, with a bright circular boundary surrounding a very dark interior. Trailing the larger CMEs were substantial outflows of small blobs and flux-rope like structures within striated ribbons, lasting for many hours.  When the  heliophysics plasma sheet (HPS) was inclined, as it was during the days around perihelion on November 21, 2021, the outflow was over a very wide latitude range.  One CME was overtaken by a faster one, with a resultant compression of the rear of the leading CME and an unusual expansion in the trailing CME. The small Thomson Surface creates brightness variations of structures as they pass through the field of view. In addition to this dynamic activity, a brightness band from excess dust along the orbit of asteroid/comet 3200 Phaethon is also seen for several days.

\end{abstract}


\section{Introduction} \label{sec:intro}
Space observations of the K-corona (the scattering of photospheric light by the free electrons in the corona) began in 1971, when the first white light coronagraph \citep[][]{Koomen1975} was launched on board the Orbiting Solar Observatory Number 7 \citep[OSO-7;][]{Follett1974} spacecraft (S/C). This was followed by a series of five missions \citep[][]{Howard2006}.  They were all at 1 au ($\sim$215~R$_\sun$) and also all along the Sun-Earth line except for the Solar Terrestrial Relationships Observatory \cite[STEREO;][]{Kaiser2008} mission.  STEREO consisted of two identical S/C drifting at the rate of about 22.5\degree~per year from Earth, one leading Earth and the other trailing.  These missions established the significance of Coronal Mass Ejections (CMEs) to space weather \citep[e.g.,][]{Gosling1997,Schrijver2015, NSTC2015}. \\

With the launch of the Parker Solar Probe \citep [PSP;][]{Fox2016} in 2018, the study of the solar corona entered a new phase with coronal observations from locations much closer to the Sun than had ever been achieved and, significantly, from below the Alfv\'{e}n surface \citep[][]{Kasper2021}. The white light coronagraphs and heliospheric imagers observe all emissions in their wide spectral bands, including the F-corona (scattering of photospheric light from interplanetary dust grains orbiting the Sun), dust particles near the S/C, planets, comets, asteroids and galactic sources. The F-corona is significantly brighter than the K-corona at distances greater than 10 R$_\sun$. Previously, the F-corona had been sampled as close as 0.3 au by the Zodiacal Light Photometer \citep[][]{Leinert1975_ZLE} on the Helios mission \citep[][]{Porsche1981_HeliosMission}. \\ 
 
 PSP observations from the Wide-field Imager for Solar Probe \citep[WISPR;][]{Vourlidas2016WISPR} have already provided significant insights on the solar environment. Most recently, \cite{Stenborg2022} showed that a dust depletion zone exists, beginning at $\sim$19~R$_\sun$ and ending at $\sim$5~R$_\sun$, at which point the dust free zone begins. Other significant observations using WISPR have been the observations of the surface of Venus \citep[][]{Wood2022}, of dust in the orbit of comet/asteroid 3200 Phaethon \citep[][]{Battams2020}, of excess dust along the orbit of Venus \citep[][]{Stenborg2021Venus},  of CMEs \citep[][]{Rouillard2020,Hess2020,2021A&A...650A..32L,Wood2021,Braga2021} and of fine-scale structure in coronal streamers \citep{Poirier2020}.\\

On 2021 November 21 in Orbit 10, PSP reached a perihelion distance of 13.29~R$_\sun$ (94\% of the distance from Earth's orbit to the Sun). Recently, \cite{Kasper2021} reported observations of the solar wind during the perihelion pass of Orbit 8 (perihelion on April 29, 2021), which indicated that PSP had crossed the Alfv\'{e}nic critical surface at $19.7~R_\sun$ and hence entered into a stream of solar wind where the magnetic energy dominated both the ionic and electron energy. These observations confirmed the existence of the sub-Alfv\'{e}nic corona and that it extended out to a heliocentric distance on the order of $\sim 20~R_\sun$.  Thus, WISPR is embedded 'within' the Alfv\'{e}nic corona.
Imaging from such a unique vantage point presents new ways of studying coronal structures but also creates challenges for our analysis and interpretation approaches, that have been fine-tuned by decades of 1~au observations. It will require time to fully resolve the challenges.\\

We aim to provide an overview of our first impressions of the observations from this unprecedented viewpoint during the 10-day period encompassing Encounter 10 (E10, hereafter).  We are struck by the very different views being recorded and revealing more subtle processes than we are used to seeing from 1 au. We do not intend to present detailed analyses of the events here. Instead, we hope to illustrate to the community the extent and richness of the differences, which must make the comparison of solar events to their terrestrial responses more difficult. All E10 observations of CMEs show a magnetic flux rope (MFR) structure within the events \citep[e.g.][]{Hundhausen1999, chen1997,Vourlidas2013}. Some of the CMEs were simple and some quite complex such as one which ran into a slower CME ahead of it revealing details of the interaction that have not been seen before, though certainly CMEs have been known to overtake and collide with each other \citep[e.g.,][]{Gopal2001}, but the details of the interactions were not revealed.\\

The views of the CMEs from so close to the Sun are showing new details. Trailing many of the CMEs were streams of outflows appearing either as discrete, small-scale structures or together with striated rays extending back toward the Sun. Post-CME outflows are quite common and have been reported before in both white light \citep[e.g.,][]{Sheeley1997,Vrsnak2009,Rouillard2010,Webb2016,Vourlidas2018} and ultraviolet \citep[e.g.,][]{Landi2012}.  The current observations of outflows from closer heliocentric distances reveal a more complex structure and seem to last longer than when observed from 1 au.  However, the streamer blowout observations \citep{Vourlidas2018} occurring at solar minimum also seemed to be accompanied by longer outflows and rays. Instances of the subtle processes being revealed include the changing perspective of coronal structures, the coronal origins of the Heliospheric Plasma Sheet (hereafter, HPS), extreme outflows of blobs, flux-ropes, rays behind CMEs, the internal structure of CMEs, and colliding CMEs. For the purposes of this paper, we will use the terms "streamer belt" and HPS interchangeably.\\

The paper is structured as follows. Section \ref{sec:obs} briefly describes the WISPR instrument, observing programs and the methodology used to enhance the visibility of the coronal transients (Section~\ref{sec:imgProc}) and to provide an overview of the coronal variability during an encounter (Section~\ref{sec:encSummary}). 
Section~\ref{sec:activity} describes, in chronological order, the main events that occurred during E10. We conclude in Section~\ref{sec:sum} with a brief discussion of our assessment of these novel observations and their implications for coronal research.  In the two Appendices we provide detailed descriptions of two methods developed to augment the scientific return of WISPR images, which in particular have been used in this paper to enhance the visibility of the features displayed in the figures. \\

\section{Observations and Methodology} \label{sec:obs}
PSP moves around the Sun in a prograde motion that coincides with Venus' orbital plane. WISPR, the imaging instrument on the PSP S/C, is located on the ram side  to provide the coronal context for the structures measured by the \textit{in-situ} payload. WISPR comprises two telescopes, inner and outer (WISPR-I and WISPR-O, respectively) that record the inner heliosphere in white light. In \citet{Vourlidas2016WISPR} the WISPR-I and -O passbands were given as 490–740~nm and 475–725~nm, respectively.  However, in a recent study of the WISPR detectors response to the Venus nightglow, \citet{Wood2022} found that their sensitivity (although reduced) extended further into the red, up to 800~nm (for both telescopes). The angular fields of view (FOV) are 40\degree~and 50\degree~elongation in the radial direction for WISPR-I and -O, respectively. At the E10 minimum perihelion distance of 13.29~R$_\sun$, the 1~au resolution (2-pixel) of 2.34 arcmin becomes 8.8 arcsec for the 1~au equivalent resolution.\\ 

\begin{figure}
\centerline{
\includegraphics[scale=0.3]{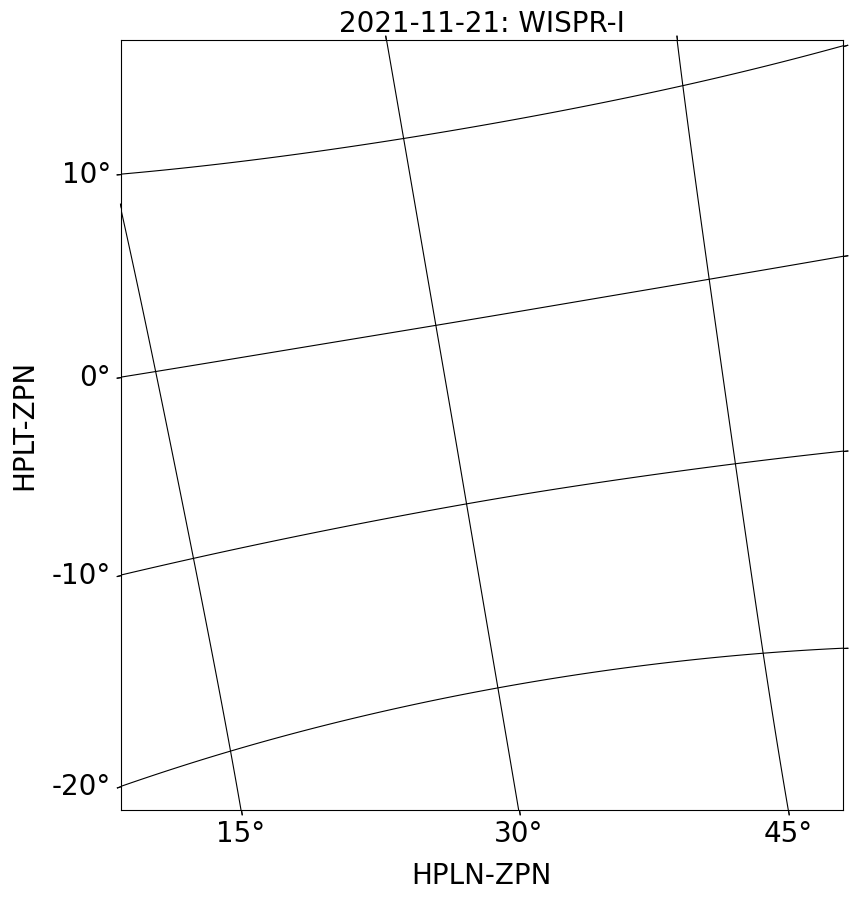}
\includegraphics[scale=0.3]{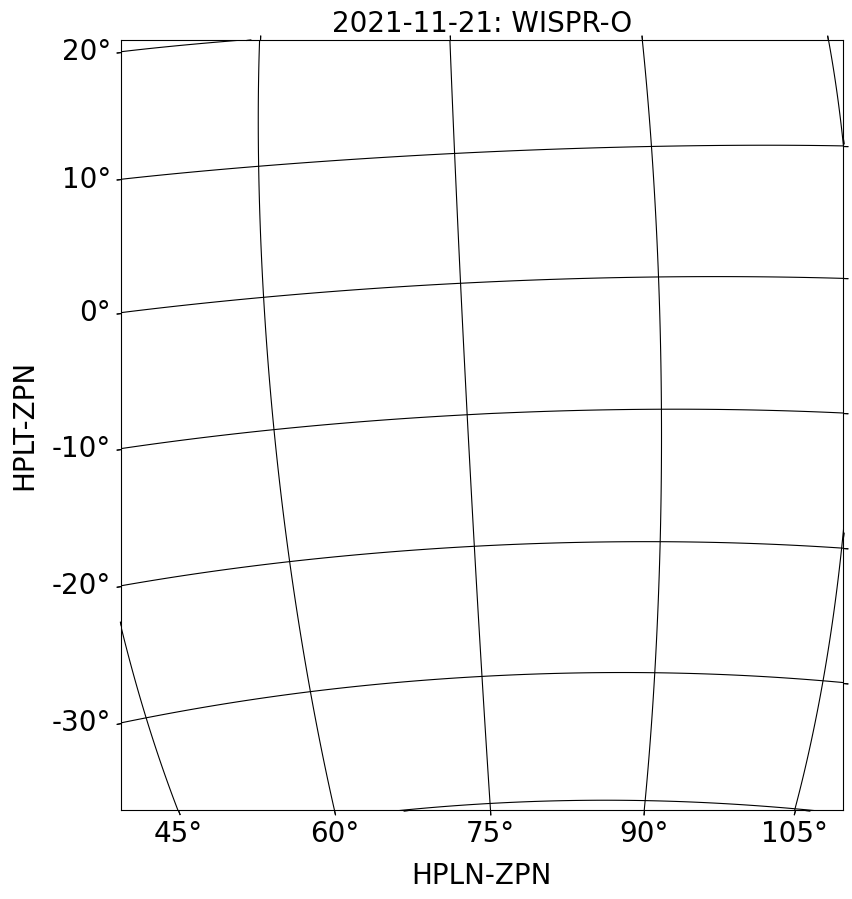}
}
\centerline{
\includegraphics[scale=0.5]{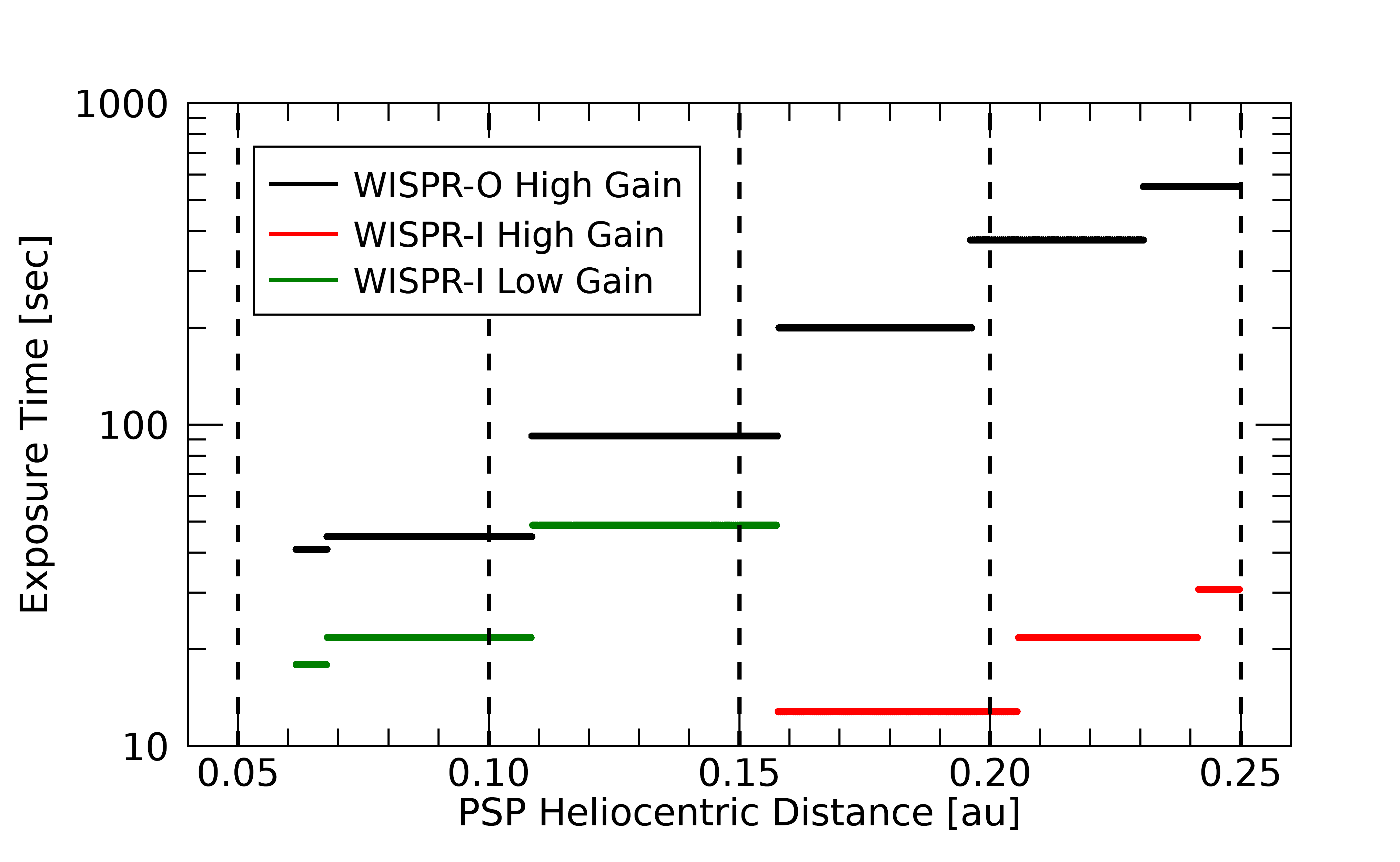}
\includegraphics[scale=0.3]{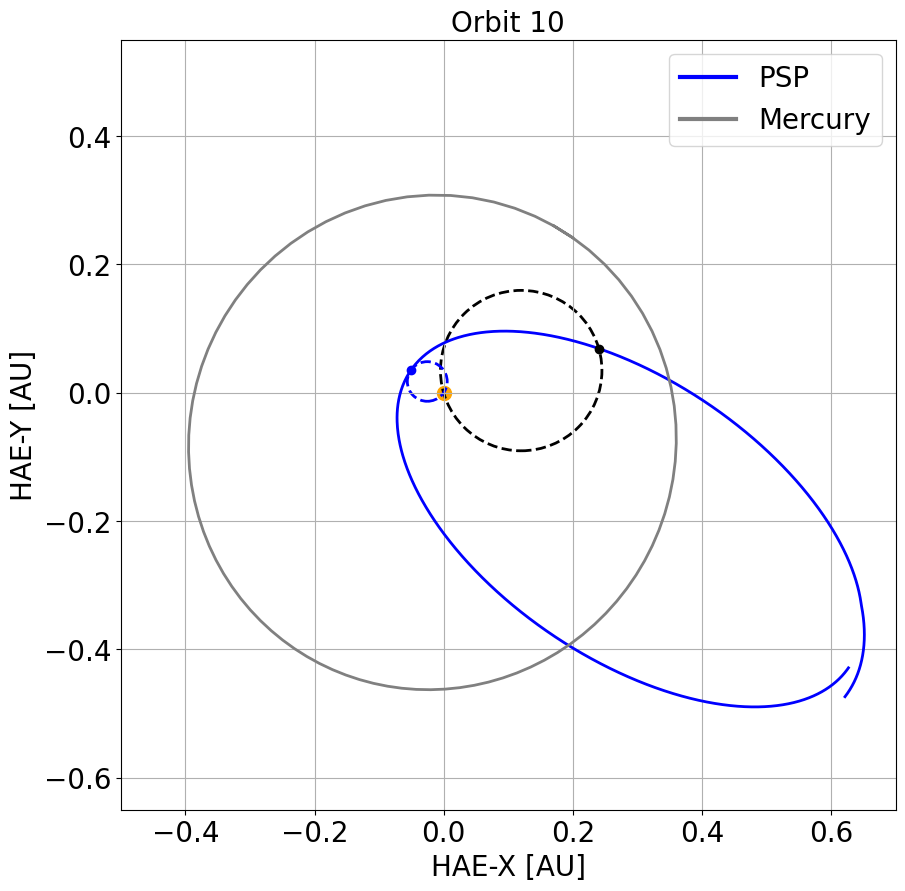}
}
\caption{\textbf{Top row:} Field of view of the WISPR telescopes. The grid depicts the fixed angular FOV in HPC coordinates for the perihelion date in Orbit 10. The HPC coordinate grid reveals the roll of each instrument frame as well as the optical distortion on the detector. \textbf{Bottom left:} Exposure Times (ET) during the synoptic observing program in Encounter 10. The ET during the inbound and outbound segments (both telescopes) of the Encounter have a similar profile. The increase of the ET for WISPR-I below 0.157 au is due to the different gain setting used for the WISPR-I images below that S/C distance. \textbf{Bottom right:} Top down view of the Thomson Surface at the start of the science encounter (black-dashed circle) and at perihelion (blue-dashed circle) of Orbit 10. The two positions of PSP are depicted by the black and blue dots, respectively. The Sun is represented by the yellow dot at [0,0]. The orbits of Mercury and PSP are marked with the black and blue continuous lines, respectively. }
\label{fig:WISPR_fovs}
\end{figure}

Figure~\ref{fig:WISPR_fovs} (top panels) illustrates the angular coverage of each telescope in the Helioprojective Cartesian (HPC) coordinate system \citep{thompson_2006, Thompson2010} when the S/C is in the science attitude, i.e., Sun-pointed with a roll angle of zero degrees. HPC coordinates are projections of Heliocentric Cartesian coordinates onto the plane of sky of the observer (the center of the system being the observer), where the x-coordinate (denoted HPLN-ZPN) is the longitude, and the y-coordinate (denoted HPLT-ZPN) is the latitude.
In this system, the Sun is at [$0^\circ$ HPLN-ZPN, $0^\circ$ HPLT-ZPN]. Note that while the angular extent of the FOV of the WISPR telescopes remains constant during the orbit, the spatial extent covered depends on the S/C heliocentric distance, in contrast to the situation for 1~au imagers.\\

The WISPR images are recorded on an Active Pixel Sensor (APS) detector ($1920\times2048$~pixel$^2$) on each telescope \citep{Korendyke2013,Vourlidas2016WISPR}, normally binned 2x2 before downlinking. The images are calibrated in units of Mean Solar Brightness \citep[MSB;][]{Hess2021}. The calibration has been very stable except for the updates to the stray-light due to diffraction as the S/C approaches the Sun. Also, no stray-light increases from dust impacts on the lenses have been detected so far. \\

The image cadence was 15 min for E10 (S/C distance below 0.25~au) and 30 min outside the lowest part of the encounter. The exposure time varies with heliocentric distance as shown in the bottom left panel of Figure~\ref{fig:WISPR_fovs}, where WISPR-O exposure times are depicted in black color and WISPR-I in green and red colors. To simplify commanding, the exposure times are changed in a stepwise fashion. The exposure increase for WISPR-I below about 0.157 au (green dots) is because the images close to perihelion are taken at the detector Low Gain setting \citep{Korendyke2013, Hess2021} rather than High Gain to increase the full well, and hence the signal-to-noise ratio (SNR), of the detector in the presence of higher signal intensity.\\

\begin{deluxetable}{lccccc}
\tablecaption{Speed and Location of PSP at the beginning, perihelion, and end of Encounter 10.}
\label{tab:params}
\tablewidth{0pt}
\tablehead{
\colhead{Property} & \colhead{Date} & \colhead{Distance (R$_\sun$)} & \colhead{Speed (km/s)} & \colhead{HAE-Long. (\degree)} & \colhead{HAE-Lat. (\degree)}
}
\startdata
Start & 11/16/2021 0418UT & 53.8 & 70.3 & 16. & -2.95 \\
Perihelion & 11/21/2021 0848UT & 13.28 & 163 & 145.50 & 3.17 \\
End & 11/26/2021 1418UT & 54.30 & 69.79 & 275.7 & -1.13\\
\enddata
\end{deluxetable}

In Table~\ref{tab:params} we show the time, S/C speed, and S/C location \citep[heliocentric distance, longitude, and latitude in the Heliocentric Ares Ecliptic coordinate system, HAE;][]{thompson_2006} at three key times during E10. Note that during the 10-day perihelion pass, PSP transited across a total of 259.7\degree~of ecliptic longitudes and 6.1\degree~of ecliptic latitudes. During this interval, the heliocentric distance to PSP is changing relatively quickly. This means that the Thomson surface \citep[TS;][]{Vourlidas_2006_TS} is also changing quickly. The TS is the locus of points along the lines of sight (LOS) where the scattering angle is a right angle, i.e. it is a sphere whose diameter is the line from the Sun center to PSP. The point along the LOS where the scattering angle is 90\degree~is the closest point to the Sun and has the maximum scattering efficiency along the LOS. Since the distance to the Sun is constantly varying, the TS itself is also constantly varying. In the bottom right panel of Figure~\ref{fig:WISPR_fovs} we illustrate the relative sizes of the TS for two points in the Encounter --in the beginning at 0.25~au (black color) and at perihelion ($\sim$0.06~au, blue color). PSP orbit 10 and Mercury's orbital path are depicted for reference with the continuous blue and black lines, respectively.\\ 

For example, during the inbound segment of the PSP orbit, the size of the TS decreases as PSP gets closer to the Sun. The distance along the LOS from the structure to the TS determines the scattering efficiency (Vourlidas et al 2016). Therefore, a change in the TS with no change in the position of the structure would introduce a change in the brightness measured by WISPR.  Of course, the structure is not stationary but is moving outward and that combined with the TS becoming smaller, will cause odd variations in the brightness of the structure simply due to its position relative to the TS.\\

\subsection{Image Processing}\label{sec:imgProc}

To enhance the visibility of the discrete K-corona structures (e.g., coronal mass ejections) as well as to unveil the dynamics of the small-scale features (e.g., outflows), we developed two customized image processing techniques to be applied depending upon the feature of interest to be highlighted.  These techniques are described more fully in the Appendix. The first technique (Appendix A) exploits the time domain to create a model, at each pixel, of the baseline brightness (5-percentile level) at each time instance. The data products resulting from this approach are referred to as ''LW''. In this approach, long-lived structures that appear as stationary in the FOV of the telescopes (e.g., streamers) are removed with the background subtraction, leaving the dynamic features. To see both quasi-stationary and dynamic structures together, we developed an alternative approach. This procedure, described in Appendix B, involves the estimation of a proxy background for each image using only the image itself. The resulting background-subtracted images are referred to as "LT".    \\

Regardless of the processing, it remains difficult to appreciate the dynamic nature of the events with still images. Movies of the encounter (as well as the whole WISPR data set) are available at the WISPR home page\footnote{http://wispr.nrl.navy.mil/wisprdata. The webpage provides links to the WISPR data in various formats. These formats include individual images with different levels of processing in FITS format: Level 1 1or L1 (raw images in digital numbers or DN), Level 2 or L2 (calibrated images in Mean Solar Brightness units or MSB), Level L2b or L2b (proxies of the background scene on each individual image), and Level 3 or L3 (calibrated images with the background removed; for details see \cite{Hess2021}. The latter data set is available also in PNG format and in the form of movies in MPG and IDL MVI formats.}. In particular, a movie of the combined FOV of the two WISPR telescopes projected onto the HPC system comprising E10 in its entirety can be found at https://wispr.nrl.navy.mil/PSP-E10-Science. In this movie, the individual frames for each telescope have been processed with the technique we called "LW" to enhance the visibility of both large- and small-scale coronal transients .   \\

\begin{figure}[h]
\centerline{
\includegraphics[scale=0.6]{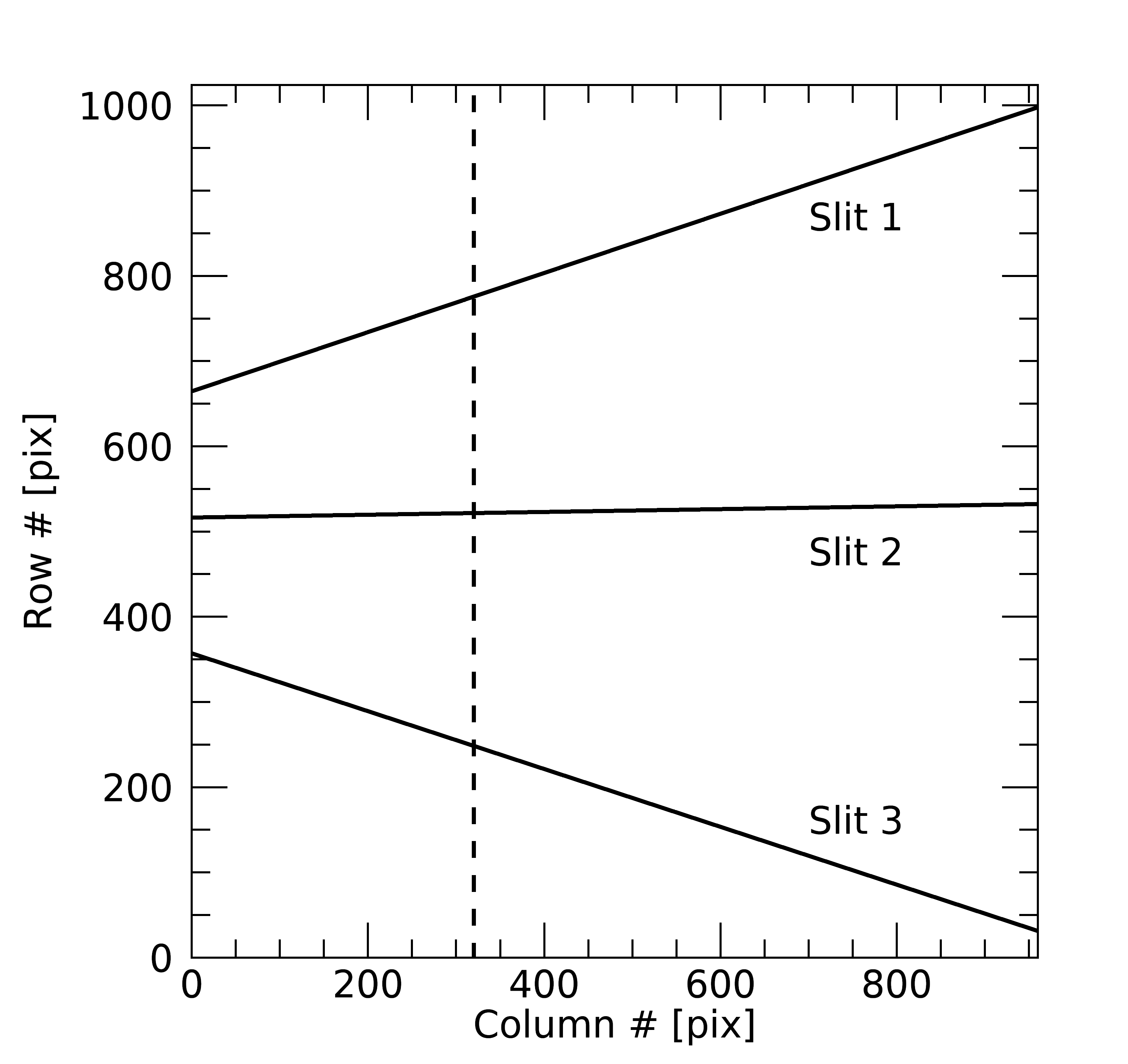}
}
\caption{Slits for the height-time and lat-time maps shown in Figure~\ref{fig:HTmaps} in a WISPR detector frame. The three radial slits were placed to encompass the FOV and capture the three regions where most of the activity was occurring (they converge at the Sun's center). The vertical slit is intended to capture the latitudinal extent of the coronal transients as well as the latitudinal evolution of pseudo-stationary structures in the FOV, e.g., streamers.}
\label{fig:slits}
\end{figure}

\subsection{Encounter 10 Summary: Height-Time and Latitude-Time Maps}\label{sec:encSummary}

A common method for visualization of radial outflows through the corona is stacking radial cuts along a given position angle versus time, originally developed by \cite{Sheeley1999_jmap} and usually refereed to as 'j-maps'. These maps capture the radial evolution but provide no information on the latitudinal evolution of the structures. The latter can be studied via time-stacking latitudinal cuts, at a given elongation. Such 'lat-maps' were first shown for WISPR by \cite{Nindos2021A&A}. Both j- and lat-maps offer a convenient method to visualize the temporal evolution of coronal structures in the radial and latitudinal directions, particularly when they are compared to the corresponding spatially resolved movies. \\

\begin{figure}[h!]
\centerline{ \includegraphics[scale=0.4]{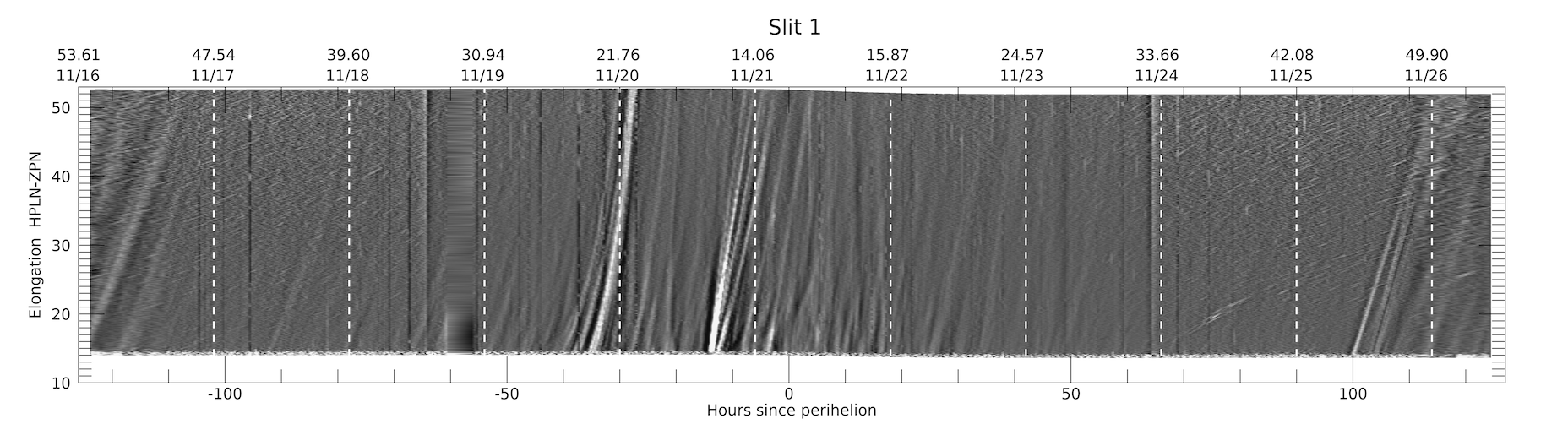} }  
\centerline{ \includegraphics[scale=0.4]{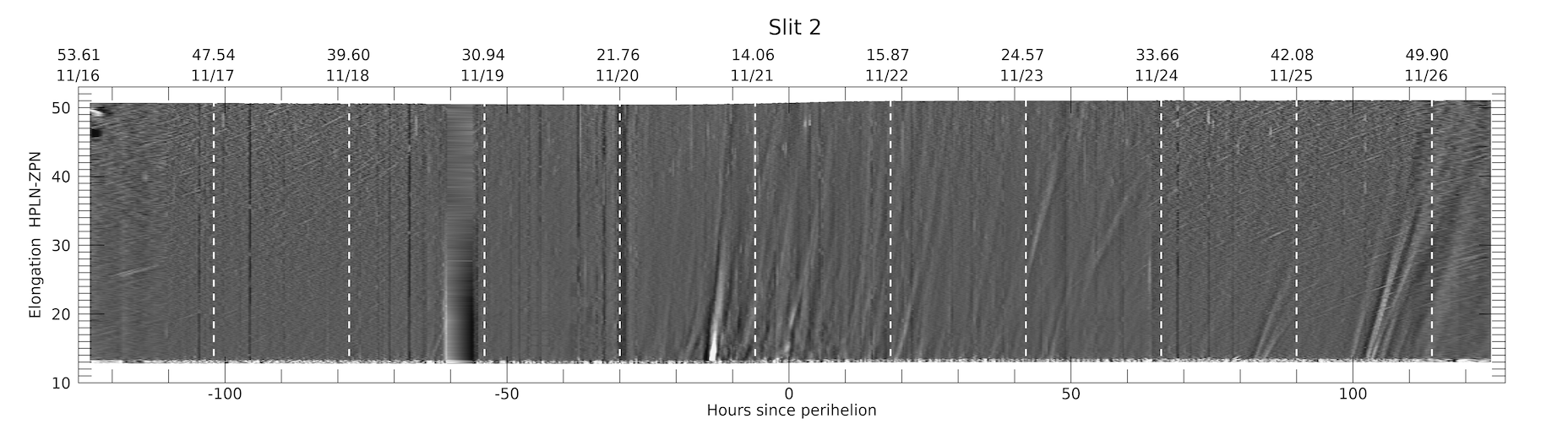} }
\centerline{ \includegraphics[scale=0.4]{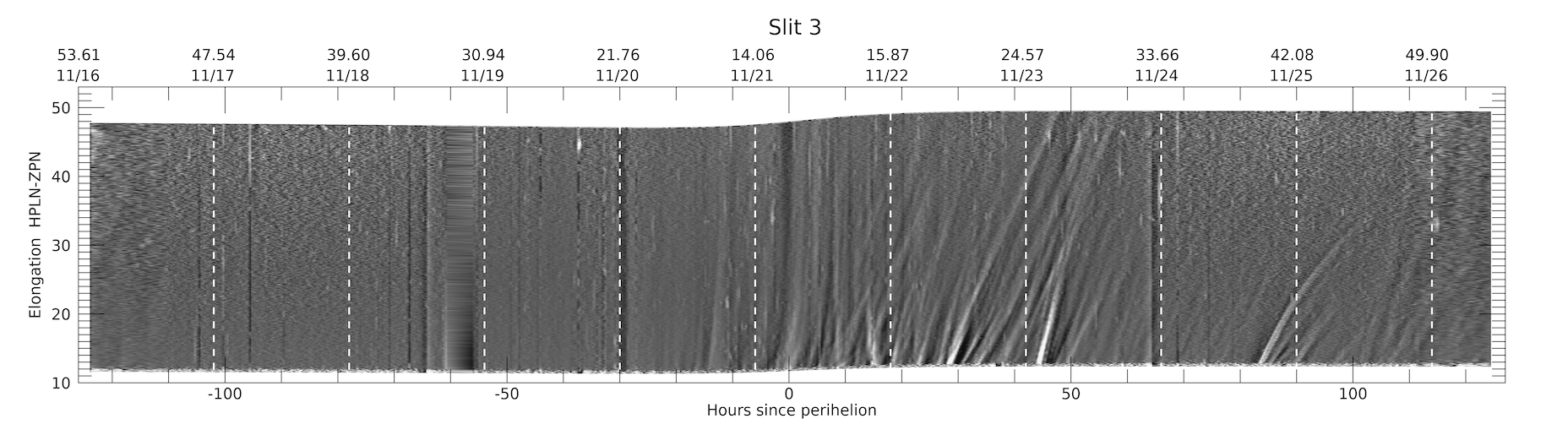} }
\centerline{ \includegraphics[scale=0.405]{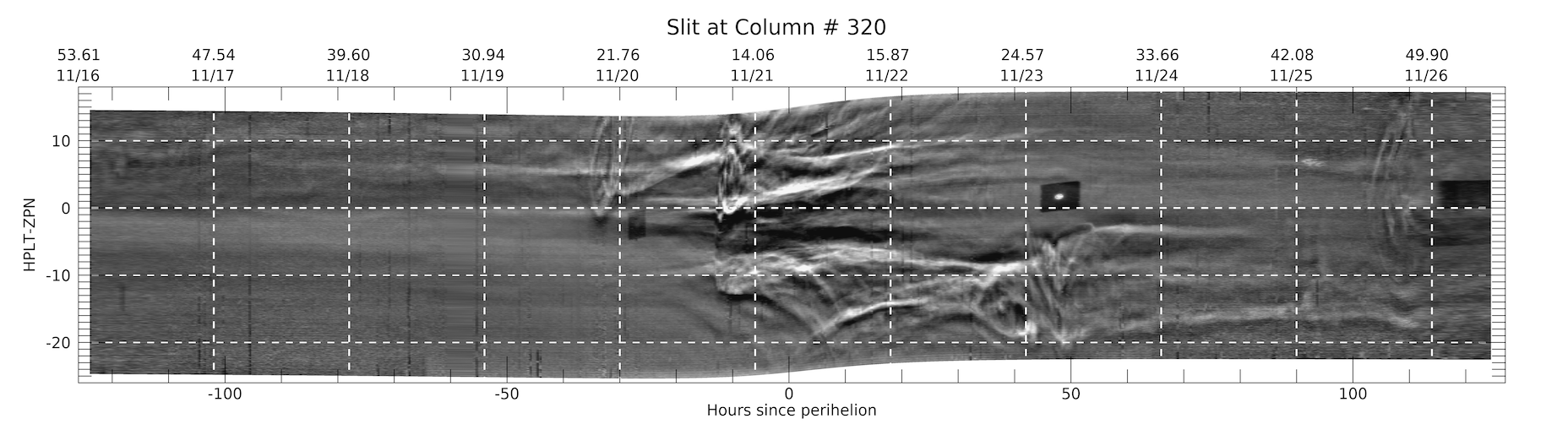} }
\caption{\textbf{Top three panels:} j-maps along the three radial directions in the WISPR-I FOV indicated in Figure~\ref{fig:slits}. The inclined features reveal the outward flow of plasma, with the curved tracks indicating the existence of acceleration (or deceleration) relative to the S/C. \textbf{Bottom panel:} Lat-map for slit at column \#320, which covers elongations (HPLN-ZPN) between $\sim$21.0\degree~and $\sim$28.2\degree~depending upon the date and position along the slit (i.e., latitude HPLT-ZPN; see Figures~\ref{fig:WISPR_fovs} --top panels and ~\ref{fig:slits}). The projection of the solar equatorial plane is at zero degrees. The start of each day (vertical dashed lines) and the corresponding S/C heliocentric distance in R$_\sun$ is indicated at the top of each panel. The wide, vertical feature at the end of November 18 is an artifact due to a 4-hr data gap in the synoptic observing program.}
\label{fig:HTmaps}
\end{figure}

Therefore, to summarize the E10 observations, we created (1) j-maps along the three radial directions and (2) a lat-map for a vertical slit as indicated in Figure~\ref{fig:slits}. The maps were made for slits within the WISPR-I FOV, where the SNR of the structures is higher, and are displayed in Figure~\ref{fig:HTmaps}. As seen in the top panels of Figure~\ref{fig:WISPR_fovs}, a given column in a WISPR image comprises several elongations depending upon (1) the specific location along the column and (2) the time of the observation. For instance, for the slit along the selected column (\#320), the elongation ranges from 21.02\degree~at the bottom to 28.81\degree~at the top when the S/C is at the start of E10, from 21.29\degree~at the bottom to 28.73\degree~at the top when the S/C is at perihelion, and from 23.27\degree~at the bottom to 27.43\degree~at the top when the S/C is at the end of E10.
The locations for the radial/latitudinal cuts were chosen to cover the  evolution of the most relevant transients (see Section~\ref{sec:activity}). The three j-maps were created from LW-processed images to enhance the dynamic features. On the other hand, the lat-map was created with LT-processed images to show both the dynamic and quiescent (streamer) features.  All available images were used and the maps were cleaned by removing the effects of the star field and of dust streaks, the latter appearing as a sudden brightening in single spots in the maps.\\

On the top of each panel, we mark the start of each day and the corresponding heliocentric distance of the S/C at the beginning of the day in solar radii. The time relative to perihelion is the horizontal axis. The patterns reveal the evolution of  coronal transients across the WISPR-I FOV.  The slopes in the j-maps reflect their projected speeds relative to the S/C; the more inclined the stripe, the faster the feature is moving. Since the stripes indicate movement relative to the S/C, their kinematics need to be considered carefully. For example, accelerating patterns may result from the approach of the S/C to the feature (virtual acceleration). This is not a common situation for 1~au observations, seen only when CMEs or stream interaction regions cross over the S/C \citep[e.g.,][]{Sheeley2008}. TS-related effects may also be visible in the maps, their signature being the appearance and/or disappearance of a given track (see, e.g., the brightest track suddenly showing up at about 25\degree~elongation on the map for Slit 2, between November 23 and 24). These effects have been predicted by \cite{Manchester2008}. Further analysis is required, however, to establish whether this is the case here. \\ 

The lat-map (Figure~\ref{fig:HTmaps}, bottom panel) is highly structured showing many features of note. Initially the map shows two faint and diffuse horizontal bands at about 10\degree~N and between 0\degree and -10\degree~S, which are the signatures of streamers seen edge on. Then, the period from the end of November 19 to the end of the encounter on November 26 shows lots of latitudinal excursions due to both coronal mass ejection (CME) events and the crossing of the streamers by PSP.  The former is characterized by sudden latitudinal excursions. For example, a first CME event occurs at the end of November 19 (see Section~\ref{sec:Nov19}). The lower part of the emission seems to form a new "streamer" at about 2-3\degree~N which moves northward to join the original streamer and both continue northward. Such signatures are also evident in the standard Carrington maps and have been used in the past to detect CMEs \citep[e.g.,][]{Hundhausen1993, Floyd2013}. Of additional interest are the features curving mainly toward the bottom of the map November 20 through November 22. Downward (and upward) curved features that reach the extremes of the FOV in the Lat-maps are caused by the rapid motion of the S/C. They imply that the observer is approaching and eventually crossing through the coronal structures (e.g., a streamer or pseudo-streamer) as discussed in \cite{Liewer2019_SCmov}.\\

When PSP's orbital speed is slower than solar rotation, the viewing situation is similar to the 1~au imaging, and the streamer belt has the familiar appearance of quasi-linear structures occasionally punctuated and/or disrupted by CMEs. For example, note the brightness signatures resembling streamer-blowout events on November 22--23 and halfway through November 25 in Figure~\ref{fig:HTmaps} (bottom) where the streamer at one latitude disappears and reforms at a different one. This latitudinal shift is caused by a disruption in the large scale magnetic field pattern. Another result of PSP's orbital speed being slower than the rotation is that the curvature of the features will be less. This was also the view in the earlier PSP orbits \citep{Poirier2020}. Instead, E10, and to some extent E8 and E9 orbits (not discussed here), show this striking transition from the 'classical' white light view of the streamer belt to a view with a multitude of fine-scale structure, while PSP super-rotates, i.e., while PSP's longitudinal speed is faster than solar rotation (for orbit 10, super-rotation occurred between November 18 19:38 UT and November 23 21:08 UT, and  when the S/C is within 0.152 au). A deeper analysis and interpretation of the structures seen in the lat-maps is currently in progress and will be reported in the near future. In the rest of this paper, we refer back to these maps as we detail the main events in E10.\\

\section{Events observed during E10}\label{sec:activity}

\subsection{2021 November 16 Event} \label{sec:nov16}

\begin{figure}[!h]
\centerline{
\includegraphics[scale=0.13]{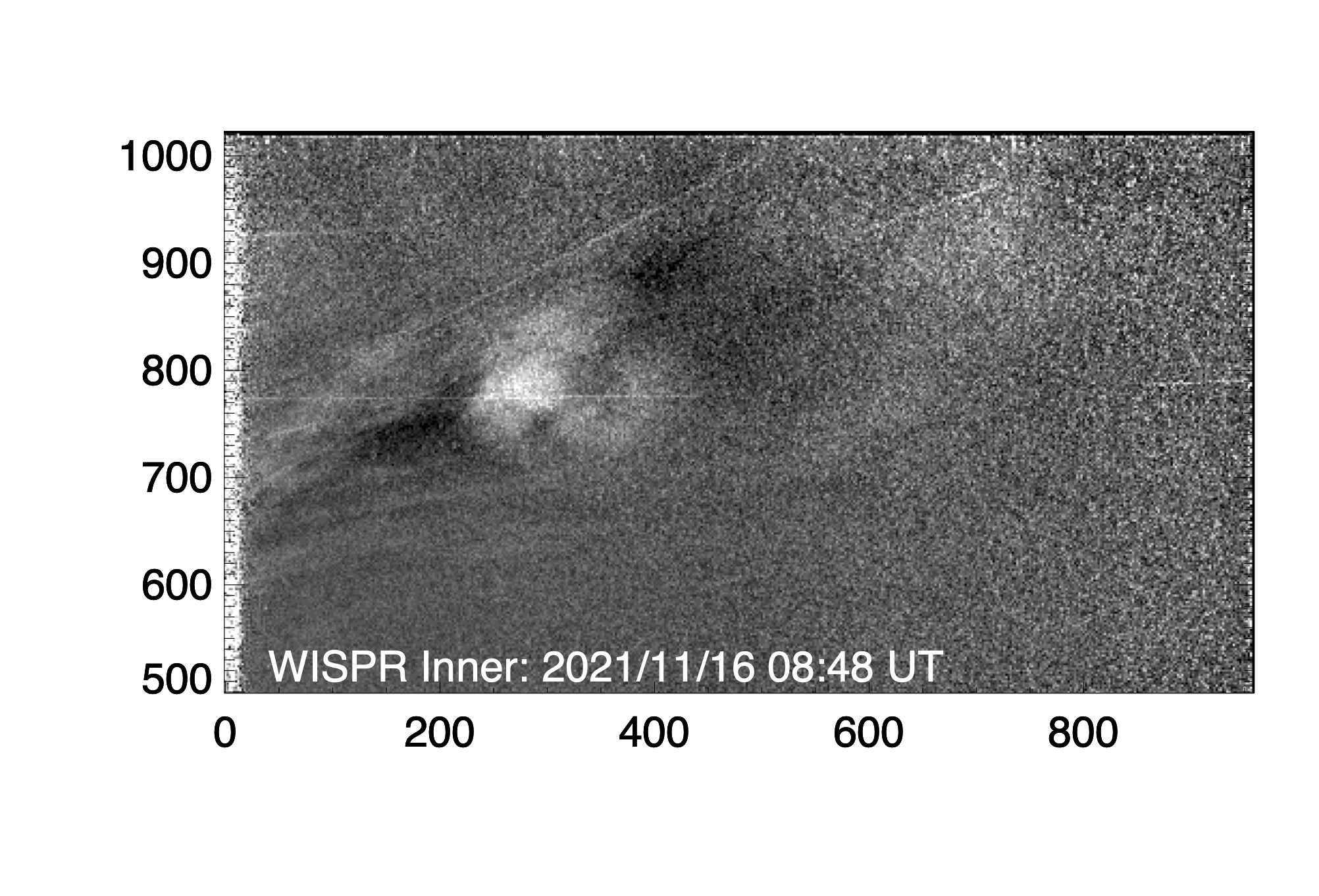}
\includegraphics[scale=0.13]{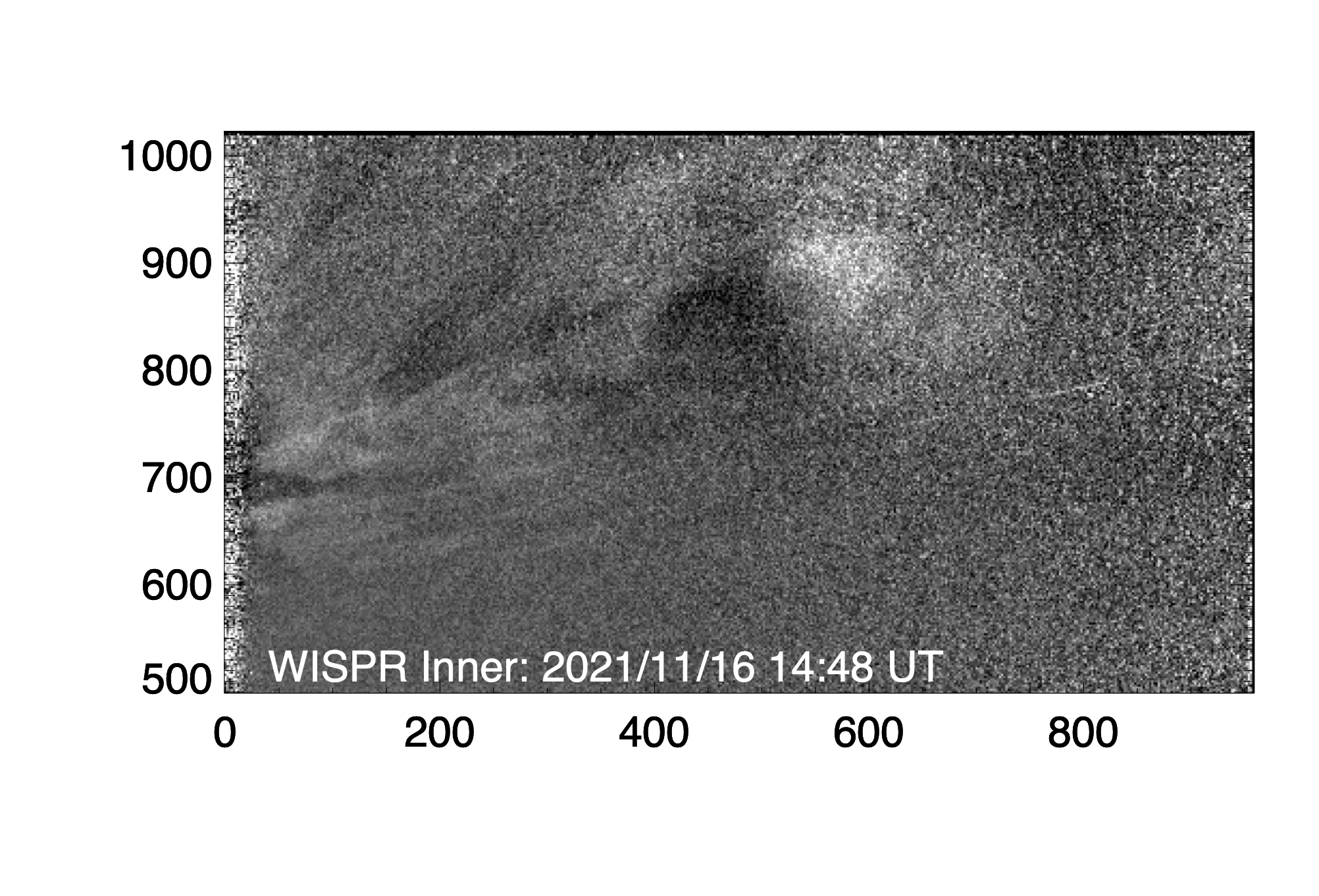}
}
\caption{Development in the upper half of the WISPR-I FOV of a small CME on 2021 November 16. The axis labels are given in pixels, and are intended to indicate explicitly the portion of the images selected for display. }
\label{fig:Nov16Event}   
\end{figure}

Figure~\ref{fig:Nov16Event} shows cropped LW-processed snapshots on November 16 at 08:48 UT and 14:48 UT illustrating the evolution of a small transient recorded by WISPR-I that develops along Slit 1 (Figure~\ref{fig:HTmaps}, top panel). In the left panel, the event appears as a MFR-type CME \citep[see][for the morphological definitions]{Vourlidas2013}, preceded by a dark area and a diffuse front. The enhanced/dark areas are indicative of the excess/depletion of electron density, respectively. Six hours later (right panel), as the event reaches the edge of the FOV a region of continuous outflow appears in its trail. Signatures of this outflow can be seen in the top panel of Figure~\ref{fig:HTmaps} (very faint and diffuse stripes following the bright stripe that depicts the transit of the main event). \\

\subsection {2021 November 19 Event} 
\label {sec:Nov19}

\begin{figure}
\centerline{
\includegraphics[scale=0.5]{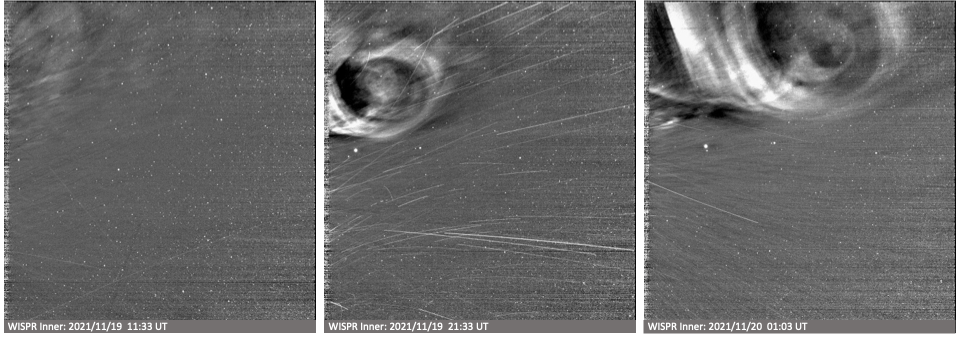}
}
\centerline{
\includegraphics[scale=0.5]{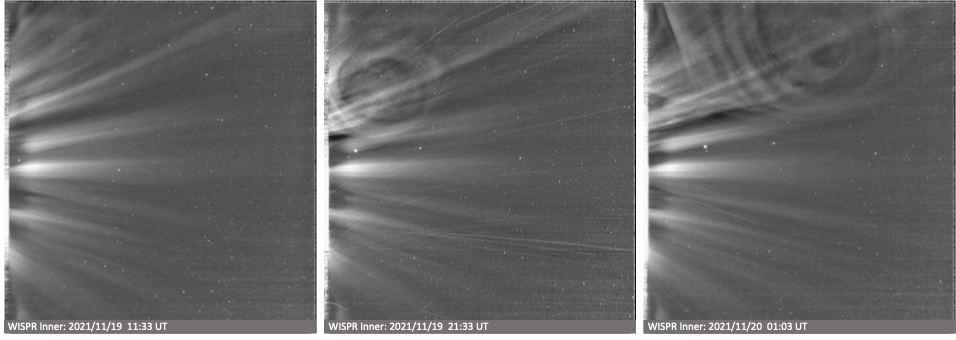}
}
\centerline{
}
\caption{Coronal activity on 2021 November 19 recorded by WISPR-I at three different time instances. \textbf{Top Row:} LW-processed snapshots. \textbf{Bottom Row:} LT-processed snapshots.
For details see Section~\ref{sec:Nov19}.}
\label{fig:Nov19Event}  
\end{figure}

Dust impacts on the S/C body are an indirect source of signal contamination for WISPR. The spallation products from these hits create bright streaks in the WISPR imagers due to solar illumination \citep{Zimmerman2021}. Their numbers increase as PSP approaches perihelion \citep[see, e.g.,][]{Malaspina2022_dust}. As a result, a larger number of WISPR images is affected at those distances. At least 30 such events were recorded on November 19. The ensuing storms of bright streaks obscured some of the images in various degrees. However, the impacts do not prevent us from capturing the fine-scale of CMEs in unprecedented detail. The CME on November 19 is a case in point.\\

\begin{figure}[h!]
\centerline{
\includegraphics[scale=0.6]{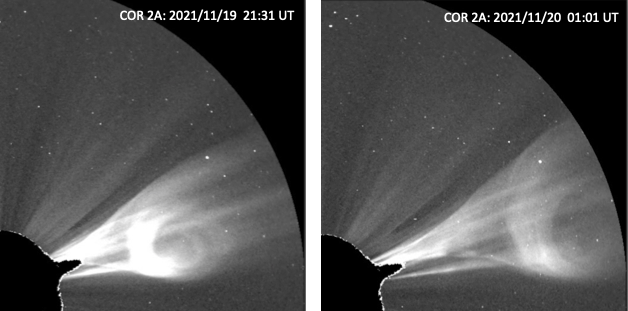}
}
\caption{STEREO-A COR2 snapshots of the November 19 event observed by WISPR (see Figure~\ref{fig:Nov19Event}). }
\label{fig:Nov19_COR2}
\end{figure}

Starting at about 03~UT, we observe a series of faint, jagged fronts along the streamer at the top edge of the WISPR-I FOV (an LW-processed  snapshot of the fronts as recorded at 11:03 UT is shown Figure~\Ref{fig:Nov19Event}, top left). The flows are overtaken by a bright front of a CME entering the FOV at 16:48~UT. Figure~\ref{fig:Nov19Event} displays two LW-processed snapshots of the CME event, on November 19 at 21:33 UT (top middle) and on November 20 at 01:03~UT (top right). Since the LW processing removes structures that appear quasi-stationary in the FOV, such as streamers, we also provide the LT-processed images in the bottom panels which reveal the dynamic CME in relation to the pre-existing streamers. Two near co-temporal snapshots of the event recorded from about 1~au by the COR2 coronagraph~ \citep{Howard2008} onboard the STEREO$-$A S/C \citep{Kaiser2008} are shown for comparison in Figure~\ref{fig:Nov19_COR2}. Both WISPR and COR2 observed the event edge-on (STEREO-A at $\sim$21\degree~and PSP at $\sim$62\degree~ecliptic longitude).\\

\begin{figure}
\centerline{
\includegraphics[scale=0.29]{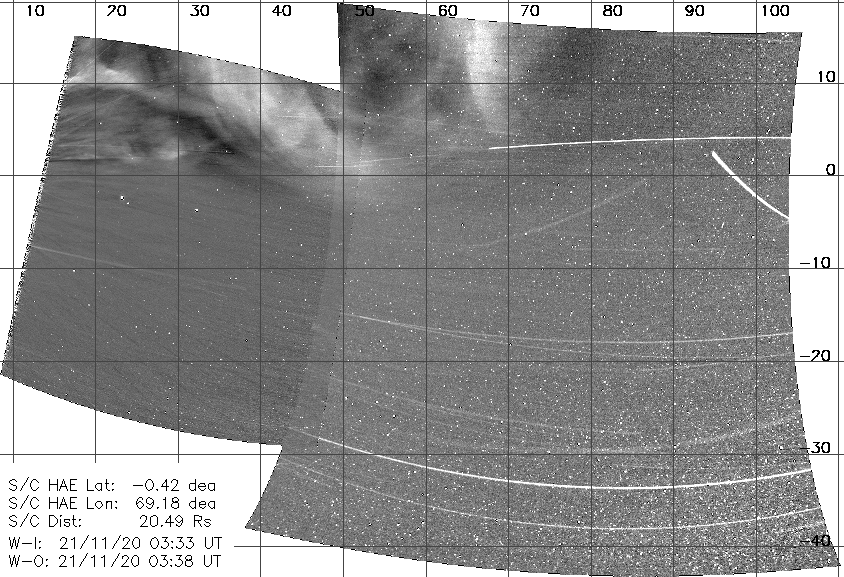}
\hspace{12pt}
\includegraphics[scale=0.29]{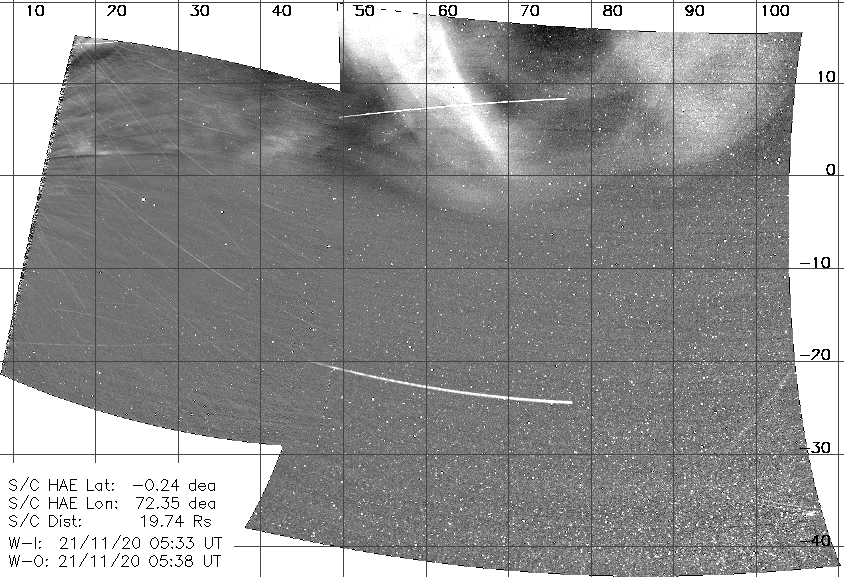}
}
\caption{WISPR-I/WISPR-O composite of the main event on 2021 November 19 at two time instances. 
The images have been projected onto the HPC grid to show the physical fields of view of both detectors together. The curvature of the images is due to a combination of the optical distortion of the lens and the spherical basis of the coordinate system. }
\label{fig:Nov19_combined}
\end{figure}

The event continues its development in WISPR-O. Figure~\ref{fig:Nov19_combined} shows two composite LW-processed snapshots of the event two and four hours later. Each snapshot combines the two WISPR telescopes in the HPC coordinate system, such that each column represents a single elongation angle and each row represents a single latitude. The back of the event exits the WISPR-O FOV on November 20 by 13 UT. The radial evolution of this transient is seen as the first bright track in the Slit 1 j-map (Figure~\ref{fig:HTmaps}, top). Its latitudinal evolution creates a rather oblong feature in the lat-map (Figure~\ref{fig:HTmaps}, bottom) showing that the CME propagates right through the center of the pre-existing streamer disrupting it, as evidenced by the lack of emission behind the CME at that latitude.  \\  

Figures~\ref{fig:Nov19Event} and \ref{fig:Nov19_combined} show a partial but detailed view of the CME that exhibits clear signatures of a MFR-type event with particularly prominent striated rays along the back of the presumed flux rope. It appears that WISPR is viewing down the MFR axis. Such events have long been observed by coronagraphs at 1~au \citep[e.g.][]{chen1997,Vourlidas2000,Vourlidas2013,Vourlidas2014}. This event is no exception. A comparison between the COR2 (see, e.g., Figure~\ref{fig:Nov19_COR2}) and WISPR images, however, shows that the WISPR images offer a much clearer view of the internal structure along with a few novel observations that raise questions for future research, as follows.

\begin{itemize}
    \item There is no compression front ahead of the CME nor was one expected since the CME projected speed is low ($\sim$375 km/s). The set of jagged fronts might have been waves ahead of the event, but they didn't appear at all like the usual wave ahead of a CME. Yet, PSP detected elevated energetic particle levels as the spacecraft approached the CME. Where/how were those particles accelerated? Could they have originated from the compression region around the MFR or were they released during the enhanced outflow period ahead of the CME?
    
    \item Contrary to what can be discerned in the COR2 images, the front in WISPR-I is not perfectly circular or symmetrically bright. Going clockwise around the CME front (middle panel of Figure~\ref{fig:Nov19Event}), we note a discontinuity across the top, an   extension across the front with a width approximately equal to the width of the preceding flows and a non-uniform brightness across the lower part. The discontinuity could be the result of the background subtraction but it is also seen in the LT image in the bottom middle panel. The lack of a corresponding feature in the COR2 images seems to indicate a LOS effect. The front extension could also be a LOS effect. We may be viewing the MFR from "underneath" or there may be an interplay between the background streamer and the MFR. The much more circular shape of the front along its southern arc seems to support the latter because it lies (apparently) outside the streamer band.\\
    
    \item Intriguingly, the central region of the MFR is not a cavity. The later WISPR images (e.g., November 20 at 01:03~UT) show another clear MFR structure lying in the approximate center of the large MFR. The COR2 image has a hint of such a structure but it is nowhere near as clear as the WISPR-I images. This central MFR appears to have its own 3-part structure (bright front, cavity, core; the latter being just a small bright region in the rear edge) and it seems to be embedded in the dark cavity of the main MFR. The LT image shows the apparent disconnect between the two MFRs best. Rare 1~au observations of such configurations exist \citep[e.g.,][]{Vourlidas2014} but the WISPR observations \citep[see also][]{Rouillard2020} indicate that their rarity may be a matter of LOS integration. With an MFR embedded in another, the white light observations would be more consistent with both the magnetic complexity measured \textit{in-situ}  \citep[see discussion in 4(b)][]{Vourlidas2019} and modeling of solar eruptions \citep[e.g.,][]{Cheng2014}.\\ 
    
    \item The axis of the FR appears to be inclined as we see a dark area off-center and we presume that the emission toward the front from that area is the forward part of the FR.  In the subsequent images the opening is enlarging rapidly, presumably because PSP is getting closer to the event.  PSP is also going underneath the FR, which causes the apparent latitude to increase during November 20, until the appearance of the next big event (see Section~\ref{sec:Nov20}). The change in the apparent latitude is also clearly seen in the Lat-time map (Figure~\ref{fig:HTmaps}, bottom).
    
    \item The bottom right two panels of Figure~\ref{fig:Nov19Event} show both the MFR and the streamers. The streamers through the MFR appear undisturbed indicating that they are most likely background structures. The streamer cannot go through the MFR --it would have been deflected otherwise, as happened in the next event (Section~\ref{sec:Nov20}). The streamer stalk along the center of the CME is visible both in the COR2 images below and in the LASCO movies, hence the structure must be over the west limb of the Sun and hence west of the CME.
  
    \item The boundary of the lower edge from the front loop to the trailing outflow is a smooth arc of enhanced emission. It is  seen as the angular ray at the back of the CME in Figure~\ref{fig:Nov19_COR2}, in the top right panel of Figure~\ref{fig:Nov19Event} and the left panel of Figure~\ref{fig:Nov19_combined}. We wonder if this is following a magnetic connection back to the Sun.\\    
    
    \item In the wake of the CME event, striated rays and post-CME outflows in the form of blobs and small FRs lasted for about a day. This is suggested in the upper left corner of both panels of Figure~\ref{fig:Nov19_combined} but is clearly seen in the movie. Also, the multiple, faint inclined ridges during November 20 in the j-map for Slit 1 in Figure~\ref{fig:HTmaps} indicate the outflows at that single angle. The next event (Section~\ref{sec:Nov20}) shows this pattern very well and so we defer a more complete discussion until then.  \\
\end{itemize}
 
In summary, the proximity of the observer to the event allowed us to observe certain features that  have not been seen so clearly before, and have given rise to more questions than answers. For instance, could the mini MFR be from prominence material or from a new flux rope emerging possibly due to reconnection or some instability?  Is the mini MFR related to a complexity at the launch or did it develop in transit?  What causes the corrugated brightness pattern along the MFR?\\

\subsection {2021 November 20-21 Event} \label{sec:Nov20}

\begin{figure}
\centerline{
\includegraphics[scale=0.5]{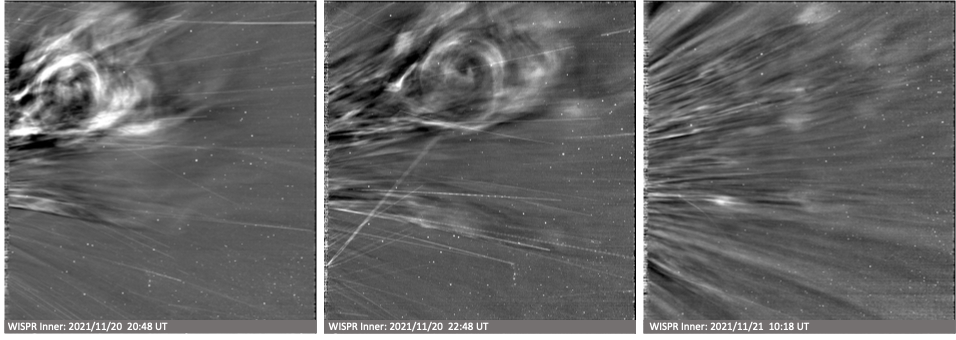}
}
\centerline{
\includegraphics[scale=0.5]{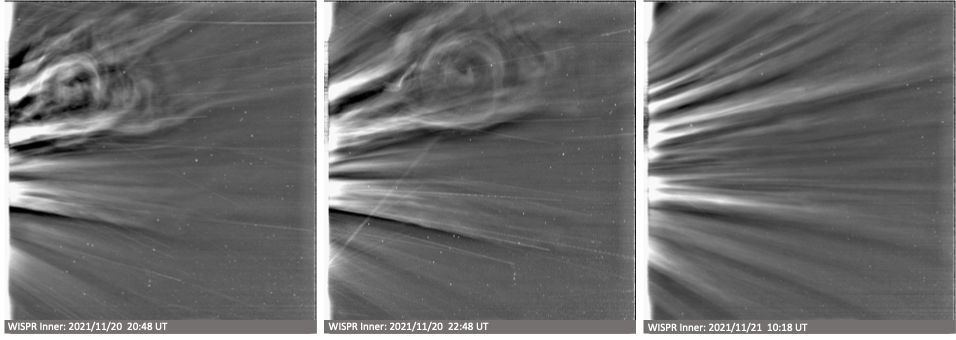}
}
\caption{Coronal activity on 2021 November 20 recorded by WISPR-I at three different time instances. \textbf{Top Row:} LW-processed snapshots. \textbf{Bottom Row:} LT-processed snapshots.
For details see Section~\ref{sec:Nov20}.}
\label{fig:Nov20Event}  
\end{figure}

Within three hours of the November 19 CME leaving the WISPR-O FOV, another CME appeared along the same position angle and with similar extent. The CME is first seen in the WISPR-I FOV on November 20 at 18:18~UT. Two time instances of the event in the WISPR-I FOV are shown in the left and middle panels of Figure~\ref{fig:Nov20Event} (LW-processed snapshots in the top row, LT-processed snapshots in the bottom row). The event differs from the previous one in exhibiting a more ragged front and streamer deflection signatures. The ragged fronts are followed by a well-defined MFR-type event, albeit with no corrugated fronts in the back. As in the previous CME, the November 20 event also shows evidence of a smaller MFR embedded within the larger MFR. The small MFR exhibits strong density depletion (in a very small region) in its center (Figure~\ref{fig:Nov20Event}, middle), indicative of strong magnetic field \citep[see. e.g., the analysis in][]{Rouillard2020}. Since the event developed while PSP was in and around perihelion, the extent of the FOV was much smaller than during the previous event and so its motion appeared to be faster than for the previous event during its early stages (see, e.g., the more inclined appearance of its track in the j-map for Slit~1 in Figure~\ref{fig:HTmaps}, top). This CME is, however, slower (between 200 km/s and 250 km/s). The latitudinal appearance of the two events, however, is very similar (Figure~\ref{fig:HTmaps}, bottom).\\

\begin{figure}[h!]
\centerline{
\includegraphics[scale=0.55]{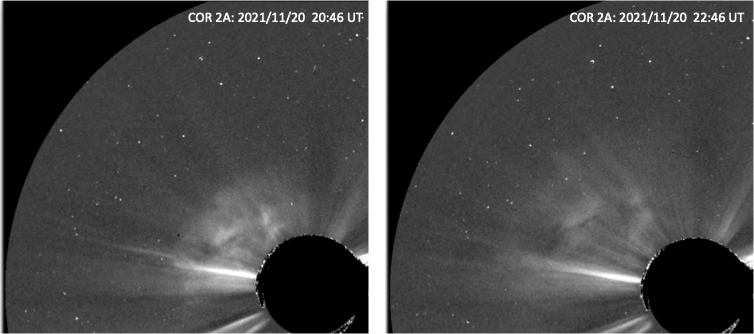}
}
\caption{STEREO-A COR2 snapshots of the November 20 event observed by WISPR (see Figure~\ref{fig:Nov20Event}).}
\label{fig:Nov20_COR2}
\end{figure}

The most striking aspect of the WISPR observations is the detection of another internal MFR-like structure in this CME like the previous event. Unlike the previous event, the MFR is interacting with the pre-existing streamers and distorting the shape, so we can't definitively identify departures from the "ideal". The COR2 observations (Figure~\ref{fig:Nov20_COR2}) are from a very different viewpoint relative to WISPR (STEREO-A at $\sim$22\degree~and PSP at $\sim$111\degree~ecliptic longitude). The COR2 view shows no such MFR structure for the main event (nor for the mini MFR), not even a front. In fact, the event appears in the COR2 images as the bulging and subsequent evacuation of the streamer, initially disrupted by the previous CME. The roundness of the outward propagating COR2 flanks is suggestive of a MFR-type eruption but the COR2 images alone would be insufficient to make a convincing case. We have been categorizing such events as 'Other'-type \citep[e.g.,][]{Vourlidas2013,Vourlidas2017} and made the point that their MFR structure could be hidden due to LOS geometry, thus confusing the discussion on whether all CMEs are the results of MFR ejection, as theory suggests \citep[e.g.,][]{Vourlidas2013,Patsourakos2020}. The WISPR observations strongly indicate (1) that, indeed, the LOS plays a major role in the apparent morphology (and interpretation) of a CME, and (2) that the number of ejected MFRs is much larger than has been realized.\\

\begin{figure}
\centerline{
\includegraphics[scale=0.29]{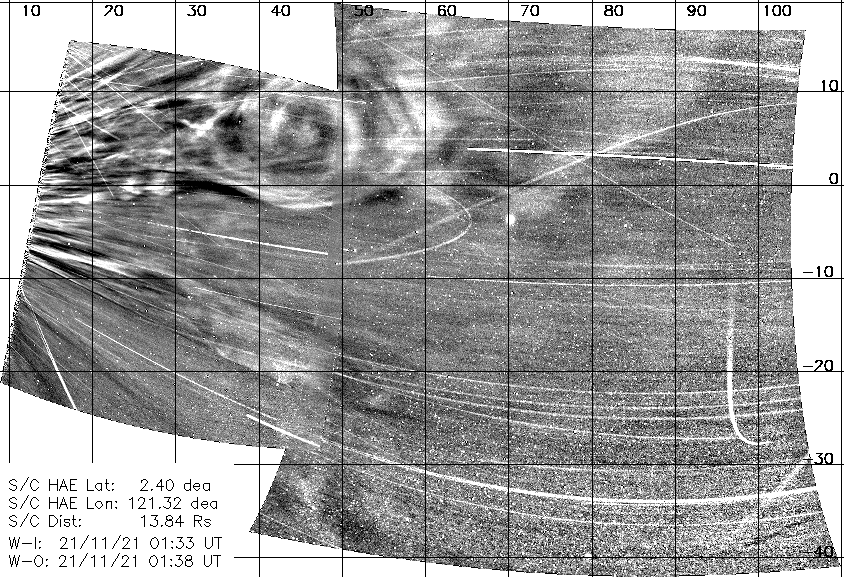}
\hspace{12pt}
\includegraphics[scale=0.29]{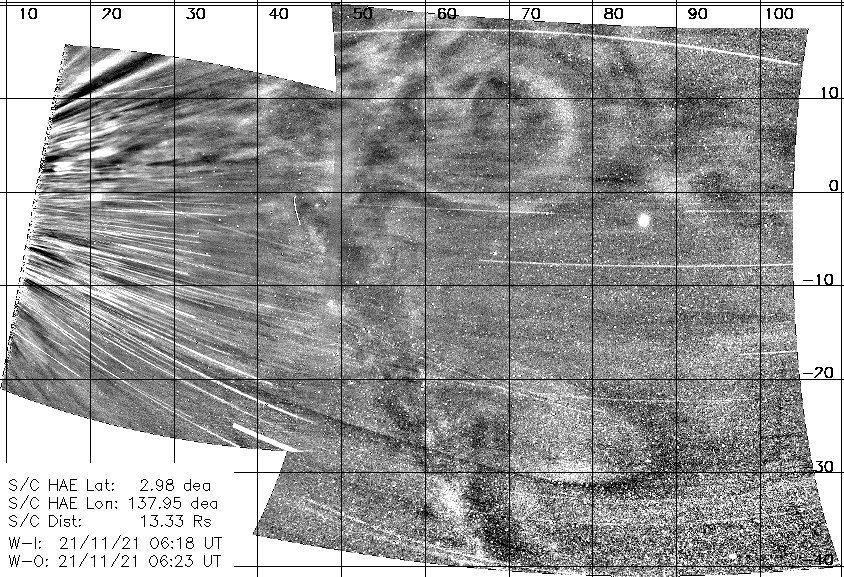}
}
\vspace{12pt}
\centerline{
\includegraphics[scale=0.29]{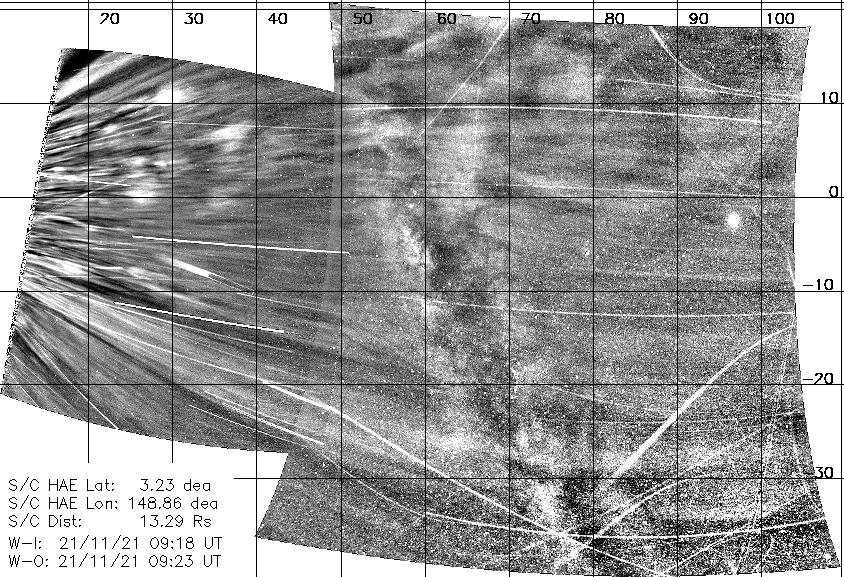}
\hspace{12pt}
\includegraphics[scale=0.29]{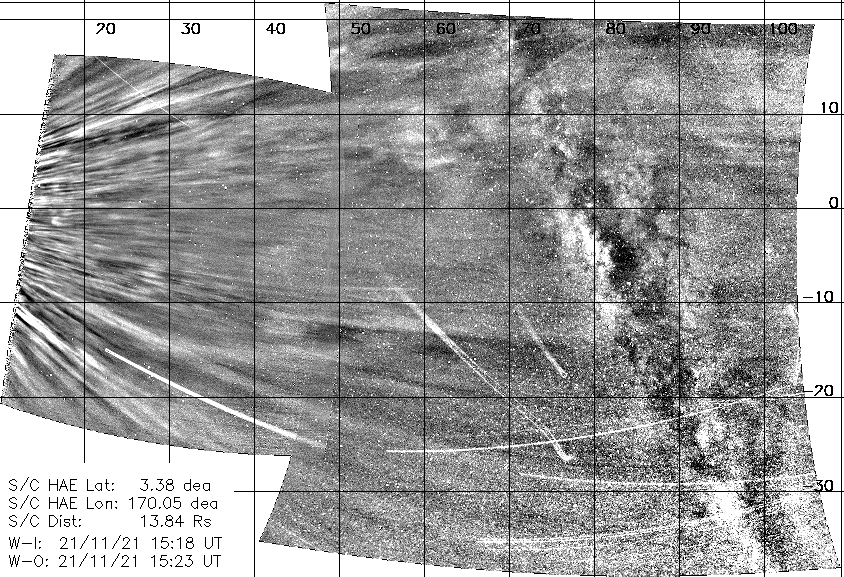}
}
\caption{WISPR-I/WISPR-O composites highlighting the coronal during activity 2021 November 20--21 (see Section~\ref{sec:Nov20}). The (LW-processed) images have been projected onto the HPC grid to show the physical fields of view of both detectors together. The Milky Way crosses the FOV of the WISPR telescopes during the development of the events. The inner telescope clearly shows the striated nature of the outflows in the HPS after the passage of the big CME event. }
\label{fig:Nov20-21_combined}
\end{figure}

Another important, and related, consideration is that the CME occurred through the PSP perihelion (which occurred on November 21 at 08:48 UT) when the S/C attained its maximum (so far) speed of 163 km/s. During the transit of the event through the WISPR FOV, the S/C traversed about 55\degree~of longitude. The angular change in the viewpoint geometry is wider than the apparent width of the CME and it could play some role in the observed evolution due to the rapidly changing TS geometry. To look for such effects, we need a comprehensive view of the brightness evolution of the event through as wide a FOV as possible. The two WISPR telescopes result in a composite FOV of >90\degree. The first three panels of Figure~\ref{fig:Nov20-21_combined} show the CME at three instances; in the mid-field (top left), just before (top right) and just after the perihelion (bottom left). The PSP heliocentric distance is almost constant through these eight hours, varying from 13.84 to 13.29 R$_\sun$. The composites are in the HPC projection. The Milky Way crosses the combined FOV during this time lapse.\\

The CME appearance between 1:33~UT and 6:18~UT changes little, mostly with the disappearance of internal structure. The circular morphology of the CME is clearly visible. However, there is a marked loss of visibility between 6:18~UT and 9:18~UT despite the unchanged heliocentric distance and a small 11\degree~change in longitude between the two snapshots. There is no apparent dynamical effect, such as an interaction with another event, that could account for the sharp change.\\

Such a rapid brightness change is not common, so we are intrigued by the idea of it being related to the geometry of the PSP orbit. The TS sphere is quite small at PSP perihelion (Figure~\ref{fig:WISPR_fovs}, bottom right). It is not so unreasonable then to surmise that even an 11\degree~change in viewing angle may be sufficient to produce such an abrupt visibility effect. This would be an intriguing result, if it holds up under scrutiny, because it may allow us to determine the LOS widths (and hence densities) of the CME substructure. Another possibility is that since PSP is moving quickly through the corona, the lines of sight comprise different views of the structure from different angles, which in turn could mean that the brightnesses could be quite different. A third possibility is a more traditional explanation. Namely, as a CME expands, its volume electron density ($cm^{-3}$) decreases by $r^{-3}$. We roughly estimate that at the perihelion distance on November 21 at 06:18~UT to 09:18~UT ($\sim13.3~R_\sun$) the distance covered in the FOV of WISPR-I (-O) is about $8~R_\sun$ ($10~R_\sun$), which would result in a substantial brightness decrease. Of course, any one or more of these possibilities could be at work.\\

We have also noticed an interesting feature in the top right image in Figure~\ref{fig:Nov20-21_combined}. An elongated, 'tongue'-like feature is seen developing along southerly latitudes ($-20\degree$~S~to $-30\degree$~S) . At earlier times, it appears as a rather amorphous expanding jet (ragged and elongated feature developing slightly to the south in Figure~\ref{fig:Nov20Event}, left and middle panels). It is interesting because it is propagating at roughly the same speed as the main CME to the north. The LW-processed image at 20:48~UT (Figure~\ref{fig:Nov20Event}, top left) shows some commonalities between the two events, e.g., a faint front seeming to connect the two, and similar propagation speeds. The LW movie of E10  shows that apparent relation better and seems to indicate that they are part of the same CME eruption, even though separated by nearly 30\degree~of latitude. The COR2 movies do show outflow associated with the CME across a wide range of position angles including the position of the 'tongue'-like event. Possibly, the intervening outflow is invisible due to the orientation of the streamer relative to the WISPR viewing geometry. On the other hand, the post-CME flow extends over a wide angular range (right panels of Figure~\ref{fig:Nov20Event} and bottom right panel in Figure~\ref{fig:Nov20-21_combined}). WISPR detects several blob-like features propagating radially but coherently across seemingly separate rays indicating a common origin. This view agrees with the COR2 observations that show a similar extent for the outflows. In other words, the CME and 'tongue'-like structure may be related. At any rate, this discussion provides another example of the importance of LOS projections in the interpretation of optically-thin emission in white light imagers. \\

Regarding the brightness variation of the 'tongue' structure as it develops across the WISPR FOV (compare the brightness change as it transits the FOV in the snapshots in Figure~\ref{fig:Nov20-21_combined}), we must consider the effect of the relative location of the feature to the TS, among other factors. We conjecture that due to the varying size of the TS as the S/C moves along its orbit, this structure approaches and crosses the TS. As a result, its brightness becomes enhanced during its approach due to the increased scattering efficiency at the TS and diminished after crossing it. However, as we pointed out above, the brightness variation could be due to either of those two factors (or even others), some or all of which might be in play. A detailed analysis, which is beyond the scope of this paper, is needed and encouraged to shed light into the reasons for the brightness decrease.\\

Finally, extensive outflows are seen following the passage of the CMEs, at all latitudes. This is consistent with WISPR viewing an inclined streamer belt face-on, a configuration created from a tilted dipolar or perhaps quadrupolar magnetic field. 
The outflows continue for most of November 21. As seen in the movie and in the lat-map (Figure~\ref{fig:HTmaps}, bottom), the bright rays move to the top and bottom of the WISPR FOV as PSP makes its fast longitudinal swing. This motion is the expected behavior when PSP approaches a structure, causing the apparent latitudinal extent to broaden. Moreover, the bottom right panel of Figure~\ref{fig:Nov20Event} (LT-processed snapshot) shows many ray-like features while its counterpart in the top right panel (LW-processed snapshot) shows plenty of small-scale activity along them. Hereafter, we will refer to the fine structure and activity in this morphological configuration simply as striations. As in the previous event, this post-CME activity is reflected in the faint, inclined ridges that show up in the j-map for Slit~1 in Figure~\ref{fig:HTmaps} (top) during November 21.\\

The pre-event streamers that are seen in the bottom left panels of Figures~\ref{fig:Nov20Event}~and ~\ref{fig:Nov19Event} are interpreted as density enhancements along the axis of the HPS. These brightness signatures have been seen before, but not so clearly. \citet{Thernisien2006} made an analysis of a streamer seen edge-on in LASCO/C2 and then face-on and found that the longitudinal density enhancements could be as much as a factor of 10.  After the passage of the WISPR events on November 19 and 20, the LW-processed snapshots (as well as the movie) show small-scale, propagating, brightness inhomogeneities that appear as blobs and small FR-like structures. Also seen in the LW processed images are many striations – rays that extend through much of the FOV with nearly linear rays of black and white patterns.  We note that the striations and the outflows both occur after the CME passage. Although outflowing structures have been seen before \citep[e.g.,][]{Sheeley1999_jmap,Rouillard2010, Viall2015_PeriodicDensityStructures, DeForest2018ApJ_Outflow}, these observations seem  unusual, i.e., there are many shapes that have a loop-like appearance which we are calling a flux rope. It is not yet clear how these inhomogeneities arise. Are they the signatures of separate flux systems? Do they reflect magnetic or density inhomogeneities in the coronal loops that formed the MFR? Could the MFR formation have occurred shortly before eruption, freezing in these differences? Are the striations small scale fluctuations of the streamer belt possibly caused by wobbling of the belt at the base which propagates outward or by magnetic pressure from the outflowing CME as first seen in the Skylab observations \citep{Gosling1974_SkylabCMEs}?  Modeling of the source region and its post-CME evolution may shed light on the big question of how CMES form \citep[see, e.g.,][for more details]{Patsourakos2020}.\\

\subsection {Activity on 2021 November 22-23} \label {sec:Nov22}	

\begin{figure}[h!]
\centerline{
\includegraphics[scale=0.25]{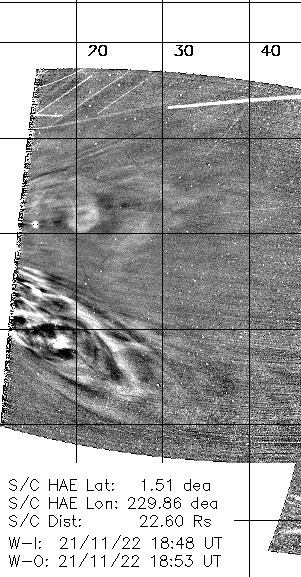}
\hspace{12pt}
\includegraphics[scale=0.25]{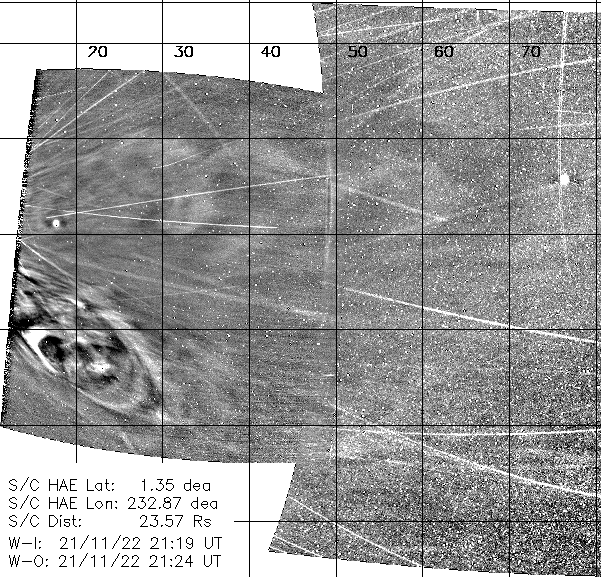}
\hspace{12pt}
\includegraphics[scale=0.25]{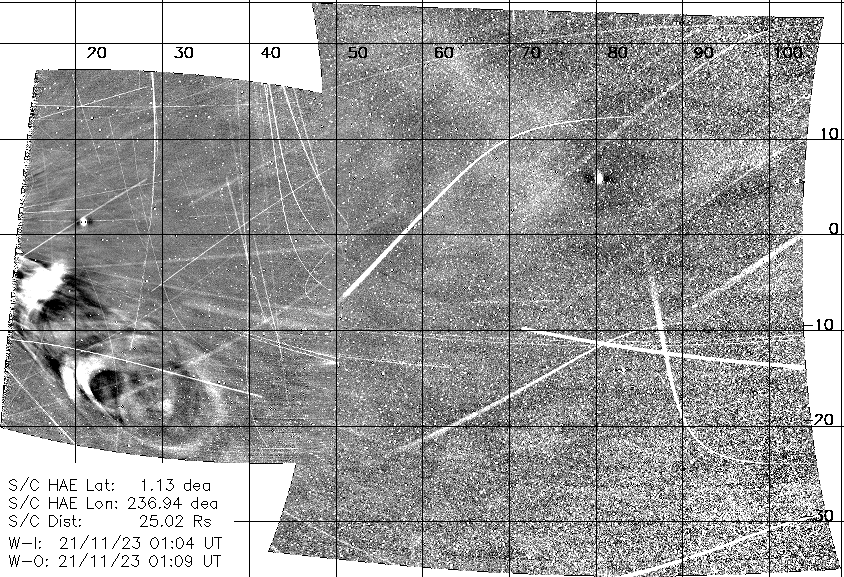}
}
\centerline{
\includegraphics[scale=0.25]{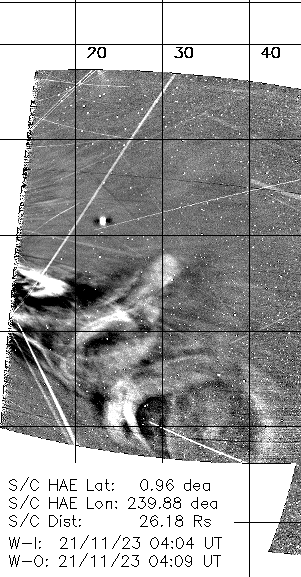}
\hspace{12pt}
\includegraphics[scale=0.25]{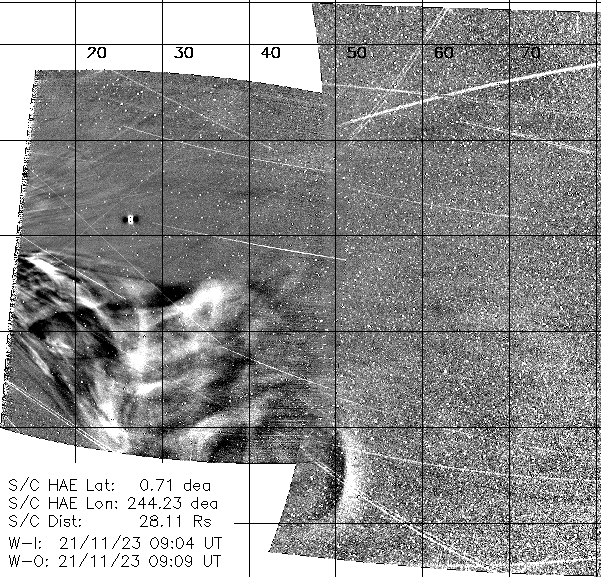}
\hspace{12pt}
\includegraphics[scale=0.25]{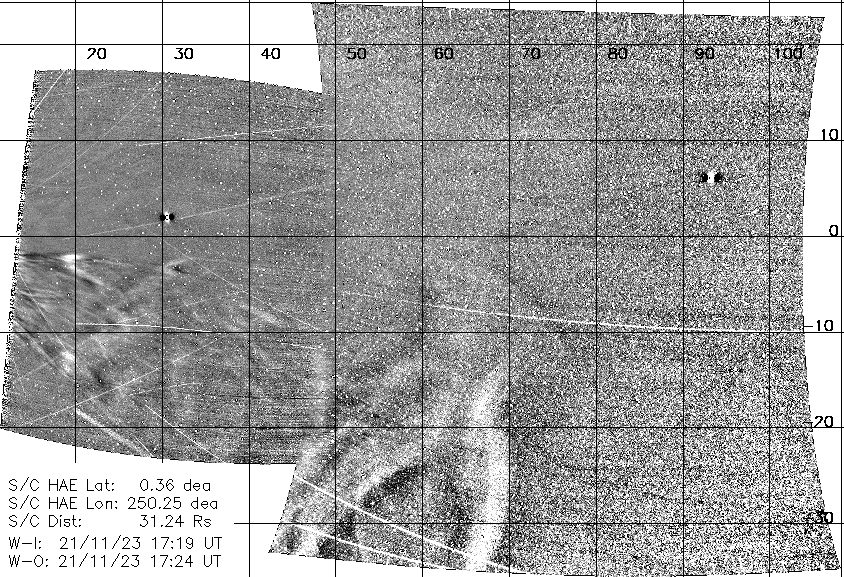}
}
\caption{WISPR-I/WISPR-O composite snapshots from November 22-23, 2021 (LW-processed) projected onto the HPC coordinate system showing the development of 1) a triple front between 0 at 10\degree~latitude (top panels), and 2) three consecutive CME events toward southern latitudes that appear to start merging before reaching the end of WISPR-I FOV (see Section~\ref{sec:Nov22}).
}
\label{fig:Nov22_23}
\end{figure}

During at least the first half of November 22, outflows continue all across the FOV of the WISPR instruments, which gradually lessen with time. The radial tracks of the outflow are visible in the j-maps shown in the top three panels of Figure~\ref{fig:HTmaps}, although they appear to be more significant at southern latitudes (see panel for Slit 3). There is continuous expansion of the apparent latitudes of the coronal rays, in which they move toward the top and bottom of the FOV of the telescopes of WISPR-O, a signature of the S/C getting close to and/or crossing the feature (Figure~\ref{fig:HTmaps}, bottom). But as the outflow lessens past midday, the rays become more linear (i.e. radial), presumably because PSP is further away from them. Outflows at northern latitudes continue past midday but to a much lesser extent. \\

Figure \ref{fig:Nov22_23} display several (LW-processed) composite snapshots, projected onto the HPC system, to show the activity during November 22 and 23. On November 22 at 12:33 UT a CME emerges at $\sim-10\degree$~ southern latitudes. Early on, the CME appears elongated, but it becomes more rounded as it evolves in the WISPR-I FOV. In addition, a diffuse structure consisting of three successive fronts is seen transiting across the middle of the image, in the corridor defined between 0\degree~and 10\degree~latitude. This event transits much faster, with the outermost front reaching the edge of the WISPR-O FOV well before the southern CME reaches the edge of the WISPR-I FOV (top right panel of Figure~\ref{fig:Nov22_23}). The faster evolution of the three-front event could be due to the direction of propagation relative to PSP. \\

Another ejection enters the WISPR-I FOV on November 23 at 00:19 UT, in the wake of the prior event (bright, small front in the top right panel of Figure~\ref{fig:Nov22_23}). The blob-like feature apparently travels faster than the preceding CME. The event becomes distorted as it evolves, due to the influence of a third event following fast behind it. The third event is first seen at 02:49 UT (bottom left panel of Figure~\ref{fig:Nov22_23}).  The  stages of this interaction are captured  in the three images in the bottom row. Some bright material (perhaps prominence material) from the third event gets pushed further to the south and slips just below bright material from the previous event.  The backend of the MFR in the first event flattens as a result of the interaction  and more bright material is pushed northward, into an unusual shape.  This shape is retained as it transits through (over?) the previous events, emerging with a distorted, wishbone shape (bottom right). The morphological changes during the interaction can also be seen in the lat-map in Figure~\ref{fig:HTmaps} (bottom). \\

\begin{figure}[!ht]
\centerline{
\includegraphics[scale=0.35]{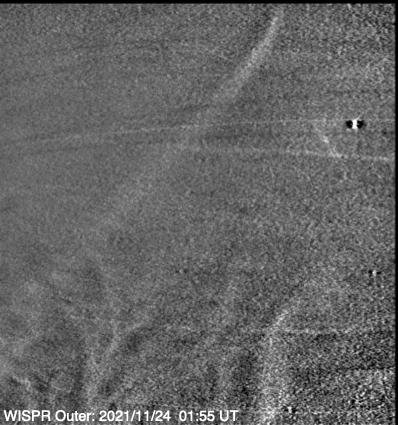}
}
\caption{LW-processed snapshot showing the dust trail associated with 3200/Phaethon in the FOV of WISPR-O. The brightness signature of the trail shows up as a diffuse band that extends from the upper right part of the image to the bottom left. The star-field has been removed from the scene with a $\sigma$-filter (as implemented in IDL Solarsoft). }
\label{fig:phaethon}
\end{figure}

Although not a coronal transient, between about November 23 at 17 UT and November 24 at 09 UT, a narrow band of enhanced brightness is visible in WISPR-O. This band, which shows up developing from the outer edge toward the inner edge of the WISPR-O FOV, appears initially as a vertical strip to become more and more curved (see, e.g. Figure~\ref{fig:phaethon}). \cite{Battams2020} identified it as a dust trail associated with the asteroid/comet 3200 Phaethon. This dust trail has been observed by WISPR in every orbit, first in WISPR-I and then in WISPR-O as the PSP orbits evolved. In a follow up study, \cite{Battams2022} finds that the dust trail is not exactly along Phaethon's orbital path, the separation between them being a function of the Phaethon's true anomaly. They point out, however, that a simple change of Phaethon's argument of periapsis by 1.0$^{\circ}$ gives a near perfect by-eye fit to the portion of the trail observed in WISPR-O.  \\

\subsection {Activity on 2021 November 24-26} \label {sec:Nov24}

\begin{figure}[h!]
\centerline{
\includegraphics[scale=0.26]{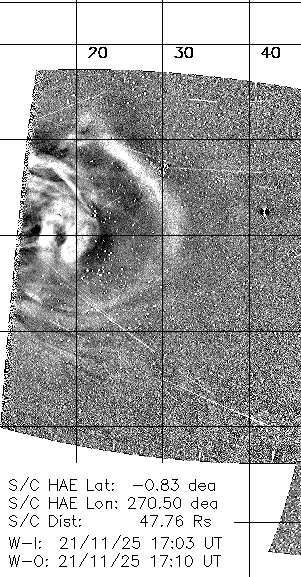}
\hspace{12pt}
\includegraphics[scale=0.26]{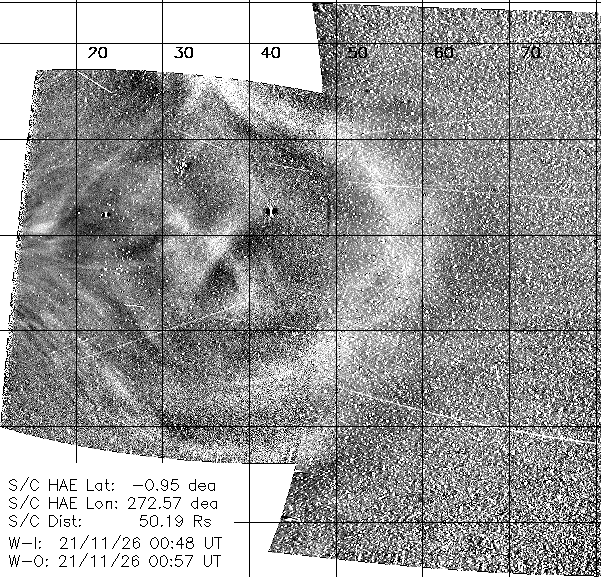}
\hspace{12pt}
\includegraphics[scale=0.26]{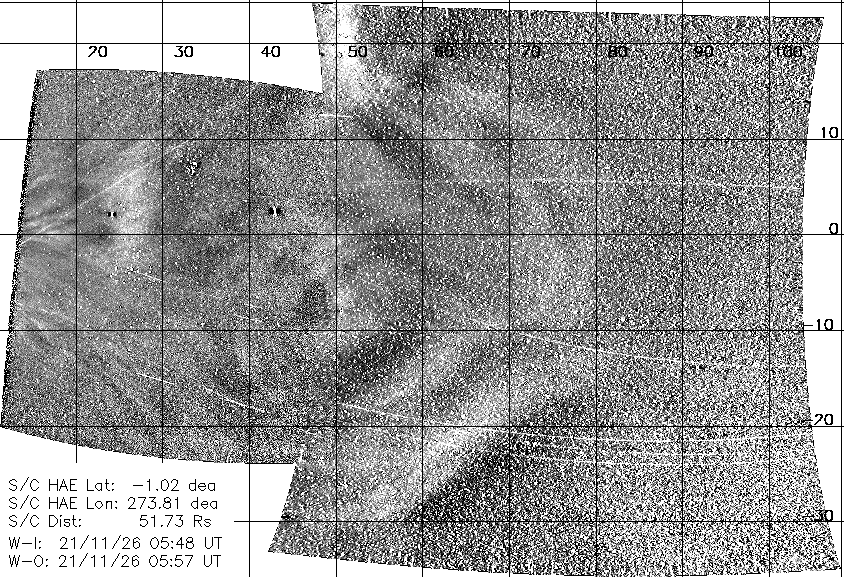}
}
\caption{WISPR-I/WISPR-O composites of three time instances of a big CME event that developed during 2021 November~25-26. Note the bright frontal arc is only about 180\degree~rather than completely around the central core of the event, indicating that the view is not edge-on. }
\label{fig:Nov25}
\end{figure}

The last event observed in E10 appears in WISPR-I on November 25 at 11:03 UT along the ecliptic, when PSP is at $\sim$45.8~R$_\sun$. A first front appears in the WISPR-I FOV on November 25 at 11:48 UT followed by another one at 14:48 UT, near the ecliptic plane. Three LW-processed snapshots of the event are shown in Figure~\ref{fig:Nov25}. The event shows up as a diffuse, broad and relatively bright and complex CME. Initially, it seems that the MFR orientation for the CME associated with each front must be inclined to the WISPR view (or the events contain no MFR) because we do not see the typical circular MFR opening surrounding a dark region. However, the movie reveals that the complex morphology seems to be the result of structure overlap as the two CME develop (the second CME moves faster and seems to merge with the previous one just before reaching the WISPR-O FOV). The faster movement of the second CME might be either a projection effect (i.e., it develops in a direction that approaches PSP more rapidly), or a result of faster development.
There seems to be prominence material inside the first CME, which appears to grow as the event develops, similar to the November 22 event.
Unfortunately, the observations of this complex event are incomplete as the WISPR program terminated as PSP exited the E10 encounter period by midday on November 26.\\

\section{Discussion and Conclusions} \label {sec:sum}

The PSP S/C achieved its closest approach to the Sun of only $\sim$13.29~R$_\sun$ or $\sim$0.062~au during the tenth orbit since launch.  This encounter occurred during the rise in solar activity, in the course of which the WISPR telescopes revealed an increase in solar eruptions, a larger amount of discrete outflows,  a "face-on" view of the HPS and several apparent crossings of coronal streamers. The nature of the E10 WISPR observations was significantly different, more complex, and more intriguing than previous encounters. This fact drove us to present these observations with minimal delay and interpretation to alert the community regarding the implications of viewing the corona (and in particular, CMEs) from 'within'. We summarize our main impressions from this exciting dataset as follows. 

\begin{itemize}
    \item All CMEs exhibited signatures of magnetic flux ropes inside them, with an orientation that enabled viewing the cross-section of the FR. Generally the MFRs retained their shape as they transited through the FOVs of WISPR-I and -O. Comparison with simultaneous observations from 1~au suggest that the detection of the MFR signatures is a matter of viewing angle/projection and not an intrinsic property of certain events. The WISPR observations strongly support the assertion (and theoretical expectation) that all CMEs are ejections of magnetic flux ropes.
    
    \item All large CMEs in E10 exhibited a complex structure in their interior. For the clearest events (Sections~\ref{sec:Nov19} and \ref{sec:Nov20}), that structure was consistent with the existence of a smaller MFR within the larger MFR. An intriguing signature of the smaller MFR was the strong density depletion at its core, which indicates the presence of strong fields \citep[see also][]{Rouillard2020}. These observations corroborate previous analyses and interpretations from 1~au \citep[e.g.,][]{Cheng2014,Vourlidas2019} and may explain the complexity of some \textit{in-situ} observations \citep[e.g.,][]{Nieves-Chinchilla2018}. The WISPR observations give new impetus to important questions on the eruption process: Has the complexity developed after the launch of the CME?  Will complexity continue to develop or will it dissipate on the way to 1 au? How would it be seen/interpreted by \textit{in-situ} measurements?
    
    \item Trailing the two large CMEs (Sections~\ref{sec:Nov19} and \ref{sec:Nov20}) was a long and, when the HPS (i.e. the streamer belt) was face-on, a very wide outflow of small, discrete blobs and some FR structures. There were also a lot of striations from the inner edge of the WISPR-I FOV to the rear of the CME, reminiscent of connections from the MFR back to the Sun. But are these rays simply density enhancements or undulations that are the result of small folds in the HPS?  WISPR was observing the HPS face-on, so the folds have to be in latitude (as opposed to our normal view of seeing the fold in longitude), and the fold has to continue smoothly over a large radial extent. Some of the rays show a curvature, which we interpret as a result of PSP being close to the structure. Are these consistent with a fold in the HPS?

    \item In one particular instance (Section~\ref{sec:Nov20}), two co-temporal, apparently unconnected CMEs, were transiting the FOV of WISPR at the same relative speed. Are they independent events or is this a result of the viewing angle and TS effects?
  
    \item In another (complex) event (Section~\ref{sec:Nov22}), an absolutely fascinating and unique interaction of three CMEs developing along the same projected direction was captured. Like all interacting CMEs of this kind, a slower CME is overtaken by faster ones following behind. But the interaction is shown in such detail that had never been seen before. A clear deformation of the trailing CMEs occurred --a flattening of their leading edge. However, in the end, they remain as two separate events as the fastest one appears to overtake and pass the other ones, keeping its new shape. This observation raises several questions: Are the three events separate ejection episodes within a larger event? Are all three connected or is this a case of overlapping but not spatially-related structures? The deformation of one the CME/ejections is dramatically obvious yet the events remain coherent; what does this imply about its magnetic coherence?
    
    \item We have pointed out several instances of brightness enhancements or decreases as the structures transit through the FOV, which we conjecture are due to the effect of the electron scattering efficiency along the LOS. The combination of the constant changing of the observer’s viewpoint with respect to a structure and the corresponding variation of the TS with respect to that structure gives an uncertainty to an interpretation. Other interpretations though are plausible, such as the decrease in the volume density of the CME due to expansion, the variation of the LOS as the S/C travels through the corona, and/or the variation of the angular subtense of the detector pixels due to the variation of S/C orbit. This will require detailed analyses beyond the scope of this paper.

    \item During the period around perihelion, the WISPR-I lat-time map shows that there is a clear apparent latitudinal excursion toward southern latitudes of the streamer belt near equatorial latitudes, an effect related to the rapid approach of the PSP to the HPS. During that time period, the number of outward flows was quite high, and encompassed a $\sim$30\degree~latitudinal width.  This is consistent with WISPR viewing the HPS face-on. The increased curvature of the corresponding rays in the undistorted images are another signature of the effect of PSP getting rapidly close to the HPS crossing.
    
    \item Finally, the dust trail nearby the orbit of 3200/Phaethon clearly appears in the FOV of the WISPR-O telescope, as it has been in every PSP orbit (Figure~\ref{fig:phaethon}).\\

\end{itemize}

We expect this paper to also serve as a demonstration of some of the image-processing techniques and tools needed for the exploitation and analysis of WISPR observations. The use of height-time and latitude-time maps along with state-of-the-art processed images helped provide convenient summaries of the dynamic nature of E10, perhaps the most dynamic encounter to date. The CME observations from WISPR showed many similarities to the 1~au observations that most of us are accustomed to. This is reassuring as it suggests that the 1~au observations (and the physical knowledge collected from them) provide reliable understanding of the CME phenomenon. In fact, the close-by views of CMEs by WISPR  appear to confirm long-held but tentative interpretations on the structure of CMEs. On the other hand, the high speed and proximity of the WISPR viewpoint to the coronal structures revealed novel patterns in the brightness, shape and dynamics of these structures that are generating new questions and enabling new methodologies for the understanding of the solar corona. \\

We invite the community to help analyze the absolutely unique data from WISPR.  PSP will have a total of 7 orbits at this same perihelion, before it performs another Venus fly-by to lower the perihelion still further to 11.4~R$_\sun$ from Sun center (five orbits with this perihelion). After a final fly-by, a perihelion distance of 9.9~R$_\sun$ will be reached. This perihelion will remain the lowest because the aphelion distance of the PSP orbit will be below the orbit of Venus.  With the rise in the solar activity cycle we expect these new orbits will reveal equally exciting observations, and combined with other space- and ground-based assets will be an outstanding dataset.  \\

\begin{acknowledgments}

Parker Solar Probe was designed, built, and is now operated by the Johns Hopkins Applied Physics Laboratory as part of NASA’s Living with a Star (LWS) program (contract NNN06AA01C). Support from the LWS management and technical team has played a critical role in the success of the Parker Solar Probe mission. The Wide-Field Imager for Parker Solar Probe (WISPR) instrument was designed, built, and is now operated by the US Naval Research Laboratory in collaboration with Johns Hopkins University/Applied Physics Laboratory, California Institute of Technology/Jet Propulsion Laboratory, University of Goettingen, Germany, Centre Spatiale de Liege, Belgium and University of Toulouse/Research Institute in Astrophysics and Planetology. G.S.and A.V. were supported by WISPR Phase-E funds. R.H. was supported by NASA grant 80NSSC19K1261. B.G., P.H., M.G.L., and N.B.R. were supported by the NASA Parker Solar Probe Program Office for the WISPR program (contract NNG11EK11I). The work of PCL was conducted at the Jet Propulsion Laboratory, California Institute of Technology under a contract from NASA. 

\end{acknowledgments}

\newpage
\appendix \label{app:Appendix}

The signal recorded in WISPR images is dominated by the F-corona/Zodiacal light component \citep[see, e.g.,][]{Stenborg2022}. Thus, in order to reveal the fainter K-corona, the ZL/F-corona (i.e., the background scene) must be removed, which is a real challenge for WISPR compared to imagers at 1~au. In particular, the high ellipticity and low perihelion of the PSP orbit (which translates into a very high speed of the S/C close to perihelion), causes a continuous change of the WISPR FOV. As one consequence of the highly elliptic orbit, far away from perihelion and aphelion, the observer's heliocentric distance is changing so rapidly that the overall F-corona brightness changes from image to image and hence the standard techniques usually employed on white-light imaging from 1~au (e.g., base or running differences/ratios) are not consistent and hence not adequate. On the other hand, near perihelion, the distance does not change that much, but the high orbital speed makes the S/C cover large heliocentric longitudes in a very short time, which makes the region of space covered by the WISPR FOV change greatly from image to image. \\

In general terms, once a background proxy is determined, the effect of the background scene can be removed by either subtracting it off or dividing by it. This creates a background difference image or a background ratio image. The advantage of the former is that it preserves the units of the calibrated data set (MSB), whereas the latter is a kind of flat-fielded version in arbitrary units. Unlike the difference image approach, the ratio image approach allows faint features to be tracked up to the very end of the instrument FOV, as it eliminates the strong radial gradient of the K-corona. For morphological studies, as the one presented in this paper, the latter is preferred.\\

Therefore, given the observational constraints, two heuristic image-processing techniques to estimate the background signal were implemented to increase the scientific return of WISPR data (in addition to the standard approach utilized on WISPR released data; see, e.g., http://wispr.nrl.navy.mil). These techniques are to be applied on calibrated WISPR images. They do not preserve the photometric calibration of the images and hence are only intended for morphological studies. Examples of the resulting data products from these two customized approaches have been shown throughout the paper. Here, we detail the broad outline of both of them with their advantages and caveats so the community can take better advantage of the WISPR novel observations.\\

\section{Exploitation of the Time Domain: ''LW'' Data Products}\label{app:A}

This approach exclusively exploits the time domain to estimate the baseline background signal at each pixel location,  $B^{P=T_1}_{i,j}(t)$, at the $T_1$-percentile level in a time window of length $L_1$, regardless of the signal in neighboring pixels. The varying F-corona brightness along with the diffuse component of the K-corona brightness, bright stars, planets, and streaks affect the determination of the baseline level. Therefore, to minimize their effect we compute the baseline brightness at the $T_1$-percentile level following the procedure below.\\

Let $I_{j,k}(t)$ be the intensity at the pixel [$j,k$] and $I^{s_0}_{j,k}(t)$ the corresponding median-filtered signal computed in an interval of size $s_0$ time units centered at the time of each image, convolved with a uniform kernel of size $k_0$. To remove the effects of both the F-corona varying brightness and diffuse component of the K-corona signal, we compute $I^N_{j,k}(t) = I_{j,k}(t) / I^s_{j,k}(t)$. Then, we median-filter the normalized brightness profile $I^N_{j,k}(t)$ using a running window of length $s_1$ to minimize the salt and pepper noise. In this way, the brightness variations along each despiked profile, $Y^{s_1}_{j,k}(t)$, reflect changes in the time evolution of the K-corona signal. To determine the baseline level of these profiles, we compute the brightness at the $T_1$-percentile, hereafter $Y^{P=T_1}_{j,k}(t)$. Note that the $T_1$-percentile of each despiked, normalized profile is a number $<1$ and hence can be used to scale $I^{s_0}_{j,k}(t)$, namely: $B_{j,k}(t) = Y^{P=T_1}_{j,k}(t) * I^{s_0}_{j,k}(t)$. \\

From the brief description above, it can be seen that $B_{j,k}(t)$ can be defined as the non- (or at least slowly-) varying basal level of the combined K- plus F-corona signal in a time window of $L_1 = max([s_0,s_1])$ time units. The step that mainly affects the contrast level of the K-corona features of interest is the way in which  $I^{s_0}_{j,k}(t)$ is computed. Here, we have described a plausible way, but the reader is encouraged to develop their own approach.    \\

\section{Exploitation of Spatial Frequencies and Time Domain: ''LT'' Data Products}\label{app:B}

In the LW approach, features that appear stationary in the FOV of the instruments, get canceled in the background corrected images. This is the case, e.g., for coronal streamers, or excess dust density features as the circumsolar dust ring near Venus' orbital path \citep[][]{Stenborg2021Venus}. To preserve these features while enhancing the dynamic K-corona transients, a plausible solution is to enlarge the time window. However, due to the brightness variability of the F-corona component associated with the rapidly changing observing distance, this alternative is not straightforward.\\

One quick way to circumvent this issue, is to diminish the effect of the background scene by computing a proxy background out of each individual image. Once each image is pseudo-corrected by their individual background, the time domain can be fully exploited without being restricted to relatively short time periods, akin to the monthly background models created, e.g., for the SOHO/LASCO mission \citep[][]{Morrill2006}. \\

The LT approach, in particular, exploits the spatial scales of the background scene on each individual image to create individual background proxy images, by taking advantage of the discrete nature of either K-corona or dust density inhomogeneities as compared to the broad contribution of the smooth background component due to the ZL/F-corona signal. One way to estimate the individual background proxies is to use an iterative recursive smoothing procedure on each individual image. This can be achieved by recursively applying a low-pass filter, e.g., by convolving each image with an appropriate two-dimensional, isotropic kernel, $K$, until the intensity gradient variations along any direction in the image are of the order of the largest scale feature. This proxy background is thus a conveniently smoothed version of the original image, where only the most extreme low spatial frequency components remain. The number of iterations needed ($n$) varies with the type and size of the kernel. The drawback of this approach is the large number of iterations needed to get the proxy for each individual image. As only the lowest frequency is of interest for the determination of the background, a faster way to determine it, is achieved by choosing a dyadic lattice for the iterative procedure. In this way $n$ reduces to $N_{max} = floor\{ [log(Z/K_z)] / log(2) \}$, $Z$ being the size of the smaller dimension of the image and $K_z$ the linear size of the kernel. The type (e.g., a gaussian, a $B_3$ spline, or a uniform kernel) and size of $K$ will influence how much of the large scale structures will appear as remnant signal in the background model.   \\

The iterative convolution with a given kernel is greatly affected by bright point-like sources like stars and planets. Therefore, the first step is to reduce/eliminate the brightness of the star field and planets in each image by applying, e.g., a sigma filter. For this paper we utilized this approach as implemented in IDL Solarsoft ("sigma\_filter.pro").  The choice of $K$,  $K_z$, and $n$ ($\leq N_{max}$) is the user's choice, and will depend upon the scale of the structures to preserve or highlight. \\

Once the proxy backgrounds are created, the varying F-corona signal is no longer an issue, and therefore standard techniques can be further applied (e.g., ''running'' or ''base'' ratio/difference, etc). The ''running'' approach suffers from the introduction of the scene of a prior frame, making the interpretation of the features observed more difficult. Therefore a ''base'' approach is preferred. An appropriate base image can be simply created by computing the T-percentile of the brightness profile at each pixel location in the time domain, the length, $L$, of the time window being a free parameter. If a short time window $L$ is considered, structures that appear pseudo-stationary will contaminate the base background and hence will introduce dark features in the resulting base ratio/difference image. Therefore, a long time window is recommended, e.g., comprising the whole encounter or, even better, comprising encounters from different orbits.    \\

\bibliography{Overvieworb10}{}
\bibliographystyle{aasjournal}

\end{document}